\documentclass[pdflatex,sn-mathphys-num%,iicol
]{sn-jnl}% Math and Physical Sciences Numbered Reference Style
\usepackage{graphicx}%
\usepackage{multirow}%
\usepackage{amsmath,amssymb,amsfonts}%
\usepackage{amsthm}%
\usepackage{mathrsfs}%
\usepackage[title]{appendix}%
\usepackage{xcolor}%
\usepackage{textcomp}%
\usepackage{manyfoot}%
\usepackage{booktabs}%
\usepackage{algorithm}%
\usepackage{algorithmicx}%
\usepackage{algpseudocode}%
\usepackage{listings}%
\theoremstyle{thmstyleone}%
\theoremstyle{thmstyletwo}%

\theoremstyle{thmstylethree}%

\usepackage{ulem}%for \sout
\begin{document}

\title[Article Title]{Second-Order Perturbative Correction for Neutrino Oscillation Tomography of the Earth}

\author%*[1]
{\fnm{Dmitry} \sur{Zhuridov}}\email{dmitry.zhuridov@uwr.edu.pl}

\author%[1]
{\fnm{Wiktor} \sur{Porwo\l}}\email{wiktor.porwol@uwr.edu.pl}

%\author[1,2,3]{\fnm{Third} \sur{Author}}\email{iiiauthor@gmail.com}

\affil%*[1]
{\orgdiv{Institute of Theoretical Physics}, \orgname{University of Wroc\l aw}, \orgaddress{\street{pl.~Uniwersytecki~1}, \city{Wroc\l aw}, \postcode{50-137}, %\state{State}, 
\country{Poland}}}

%\affil[2]{\orgdiv{Department}, \orgname{Organization}, \orgaddress{\street{Street}, \city{City}, \postcode{10587}, \state{State}, \country{Country}}}

\abstract{
The Earth’s interior at depths exceeding present direct probing limit, which is just one five-hundredth of its radius, remains unknown to a significant extent due to existing ambiguities in predicting the distribution of temperature, pressure and chemical composition on the basis of seismic and geophysical observations. Scanning the Earth with neutrinos can provide independent information for refinement of existing seismological models. However, neutrinos interact with medium extremely weakly. Only tiny TeV-energy component of the atmospheric neutrino flux is noticeably absorbed while traversing the Earth, which provides limited statistical data. Flux reduction for much more common neutrinos with energies of the order of 1~GeV is vanishing. However, on their way through the planet they undergo sizable flavor oscillations that depend on the electron number density, which provides a link to seismological models. Accurate calculation of the matter effect in these oscillations is required for successful neutrino oscillation tomography of the Earth besides growing power of neutrino experiments, e.\,g., KM3NeT, DUNE and Hyper-Kamiokande. We obtained relevant neutrino oscillation probabilities in the second order of perturbation theory. Our results are valid for a neutrino potential that is not necessarily symmetric along the neutrino path. This allowed us to implement a multi-shell evolution scheme to better account for the density jumps between various layers of the Earth.
}

\keywords{Neutrino oscillations, Earth tomography, perturbation theory, high-order corrections, atmospheric neutrinos, evolution matrix}



\maketitle

\section{Introduction}\label{sec1}

Present knowledge about the interior of the Earth is obtained mainly by means of seismology and geophysics (heat flux, geomagnetic variations, etc.)
~\cite{Poirier,McDonough,Masters}. 
Seismological models of the Earth utilize velocity-depth profiles and the equation of state to obtain matter density, pressure, and elastic moduli profiles. In particular, in the Preliminary Reference Earth Model (PREM)~\cite{Dziewonski:1981xy} our planet is divided into layers, i.\,e. spherically symmetric shells, which are separated by seismological discontinuities and associated with them density jumps. 
The main shells are core and mantle. Each of them is divided into sub-layers: inner and outer cores, lower mantle, transition zone consisting of three parts and four outer shells. However, the distribution of matter density in the mantle is known with an uncertainty of about 5\%, and for the core the uncertainty is essentially greater~\cite{Masters,Geller:2001ix,Jesus-Valls:2024tgd}. 
The problem is that the two velocity components of seismic waves, longitudinal (P-waves) and transverse (S-waves), depend on three independent parameters -- the two elastic Lam{\'e} moduli and density -- that in tern vary with the chemical composition, pressure, and temperature, which are not measured directly at depths that exceed approximately 12~km~\cite{Kozlovsky}. 
Moreover, a liquid phase in the core does not let S-waves through~\cite{Geller:2001ix,Bullen,Borisov:1986sm}. 
Modern seismic tomography allows to obtain detailed 2D and 3D models of the Earth's subsurface from seismic records by solving an inverse problem in geophysics and minimizing the difference between the created model and data observed. 
However, this requires extensive assumptions and extrapolations~\cite{Goos:2025rra,Deng:2023,Kozak:2023axy}.
The resulting ambiguities can be relaxed by utilizing complementary observations and techniques. 

Scanning the Earth by means of neutrinos is a new promising tool to explore its internal structure, in particular, the core since the neutrino signal is not affected by its phase state. 
This method relies on the weak interactions of neutrinos that for the most part easily traverse the entire planet. The neutrino, which is a mixture of states with definite masses, can be detected as one of three specific states, known as flavor neutrinos (electron neutrino $\nu_e$, muon neutrino $\nu_\mu$ and tau neutrino $\nu_\tau$), and probability of its detection depends on the distance traveled\footnote{This is convenient to measure the distance traveled by a flavor neutrino instead of the time passed since its velocity is very close to the speed of light both in a vacuum and in matter.}, i.\,e., neutrinos change flavor along their trajectories. 
These neutrino flavor oscillations are measurable and highly sensitive to properties of the medium through which they propagate. Hence, interactions of neutrinos with the Earth's matter not only cause absorption of some of them (attenuation of their flux), but also affect their oscillation picture (matter effect in neutrino oscillations). 
The emerging Earth tomography using neutrinos can be divided into two approaches: neutrino absorption tomography (NAT)~\cite{Placci_Zavattini,Volkova_Zatsepin,Nedyalkov:1981yy,Borisov:1986sm,Donini:2018tsg,IceCube:2025utw} and neutrino oscillation tomography (NOT)~\cite{Nicolaidis:1990jm,Ohlsson:1999um,Akhmedov:2006hb,Kelly:2021jfs,Jesus-Valls:2024tgd}. 
Although neutrinos in the atomic medium are scattered mainly by nucleons (protons and neutrons), the neutrino potential associated with the matter effect in neutrino flavor transitions is determined by the electron number density~\cite{Giunti:2007ry}.  
Thus, neutrino interaction properties lead to NAT and NOT sensitivity to number density distributions of nucleons and electrons, respectively, inside the Earth. 
The range of applicability of these methods is determined by the neutrino mean free pass and neutrino oscillation length(s) that essentially depend on the neutrino energy as we discuss in more detail in the next section.

The collection rate of the neutrino data, which can be utilized for the neutrino tomography of the Earth, is expected to increase significantly in the coming years, in particular, with anticipated runs of the Hyper-Kamiokande~\cite{Hyper-Kamiokande:2018ofw,Jesus-Valls:2024tgd} and DUNE~\cite{Kelly:2021jfs} experiments. For this reason, accuracy of theoretical predictions for the oscillated neutrino fluxes will require an improvement. According to PREM, density within the majority of the Earth's layers is far from being a constant. However, calculating the neutrino oscillation probabilities for a piecewise nonlinear potential is a nontrivial task. There are several methods to deal with it in the NOT framework. 
One way is to discretize PREM into a number of constant density layers (up to 128 bins for the specific studies in Ref.~\cite{Goos:2025rra}). Alternative methods include application of either perturbation theory~\cite{Brahmachari:2003bk,Akhmedov:2005yj,Akhmedov:2006hb} or Magnus exponential expansion~\cite{Supanitsky:2008eq}.

Our main interest in this paper is to obtain the neutrino oscillation probabilities for NOT with high accuracy for a realistic Earth density profile, in particular, to account for the effect of varying density within different Earth's layers. We perform this task in the perturbation theory approach and include the second-order contribution for the first time. 
To introduce the reader to this topic we discuss rough approximations in the next section. We present perturbative calculation of the matter effect up to the second order in section~\ref{sec:prturbative}. Next we discuss its application to a realistic Earth density profile and compare to the Magnus approach in section~\ref{sec:evolution}. Then we discuss the results in section~\ref{sec:discussion} and conclude in section~\ref{sec:conclusion}.

\section{Average density approximation: one-shell evolution}
\subsection{Simplified Earth density profiles}\label{sec:density_profiles}

\begin{figure}[tb]
	\centering
	\includegraphics[height=5cm]{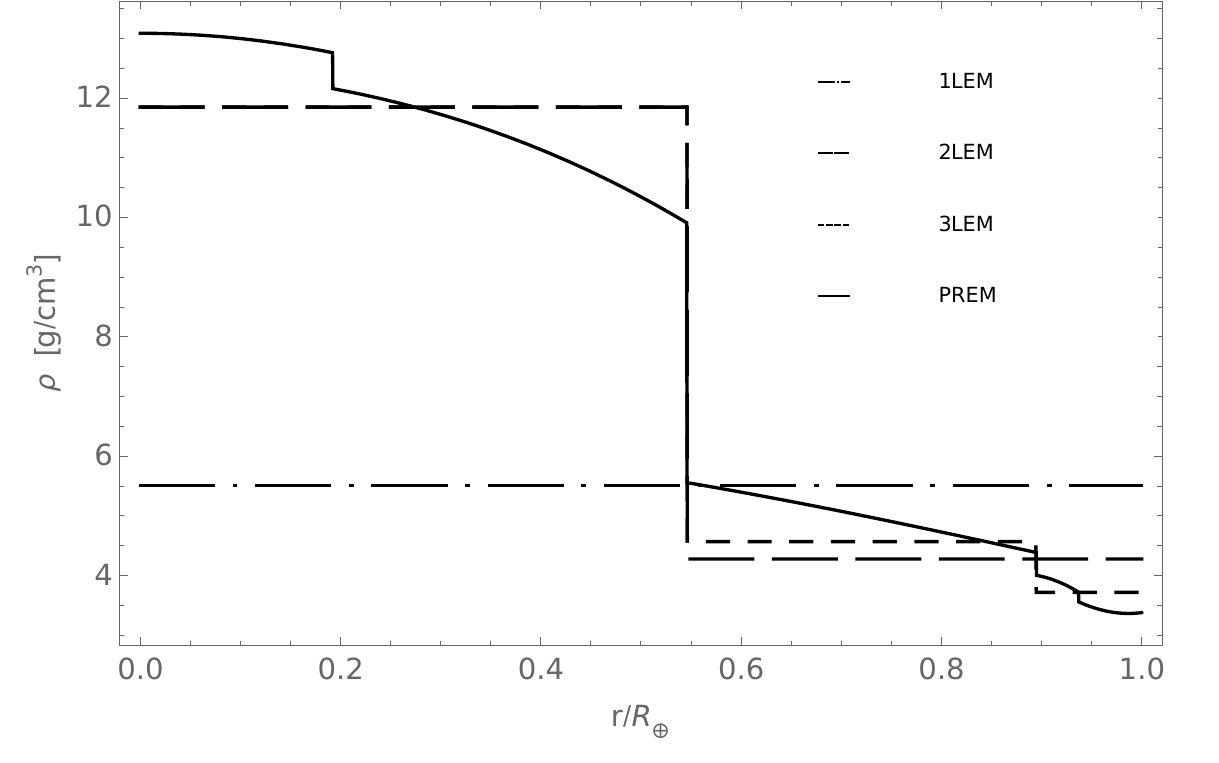}
	\caption{Matter density vs radial distance normalized to the Earth's radius for 1-, 2-, 3-layer and PREM density profiles as represented by dot-dashed, long-dashed, short-dashed and solid lines, respectively.}
	\label{Fig:rho-r}
\end{figure}

Taking into account that precision of the neutrino tomography is still far from the seismic one, the Earth's interior is often approximated by several layers of constant density, which essentially simplifies calculation of the neutrino oscillation probabilities. In particular, the following cases are considered in Ref.~\cite{Kelly:2021jfs}, which both ignore the relatively thin Earth's crust: 
\begin{itemize}
	\item {\it Two-layer Earth model} (2LEM): the core and the mantle, with densities $\rho_{\rm C}$ and $\rho_{\rm M}$, respectively, corresponding to the radial distances $r<R_{\rm C}$ and $R_{\rm C}<r<R_\oplus$, where $\rho_{\rm C}=11.85$~g/cm$^3$, $\rho_{\rm M}=4.28$~g/cm$^3$,\footnote{The value of $\rho_{\rm M}$ is fixed by the Earth mass.} %$\rho_{\rm M}=4.14$~g/cm$^3$ {\color{red}(correct $\rho_{02}$ is for $\rho_{\rm M}=4.28$)}, 
	$R_{\rm C}=3480$~km and $R_\oplus=6371$~km (upper mantle).
	\item {\it Three-layer Earth model} (3LEM): the mantle is subdivided into the lower ($R_{\rm C}<r<R_{\rm LM}$) and upper ($R_{\rm LM}<r<R_\oplus$) parts with densities $\rho_{\rm LM}$ and $\rho_{\rm UM}$, respectively, where $\rho_{LM}=4.57$~g/cm$^3$, $\rho_{\rm UM}=3.72$~g/cm$^3$ and $R_{\rm LM}=5700$~km.
\end{itemize}
For completeness we include also a trivial {\it one-layer Earth model} (1LEM), in which the Earth's density is approximated by the constant value of $\rho_0 = 5.52$~g/cm$^3$. %$\rho_0 = 5.515$~g/cm$^3$. 
The discussed Earth density profiles are shown in Fig.~\ref{Fig:rho-r}. 

\subsection{Characteristic lengths in the neutrino tomography}

The mean free pass $\ell_\nu$ of a neutrino in a medium and neutrino oscillation length(s) $L^{\rm osc}$ are important distance scales in the neutrino tomography. 
The first one can be estimated by order of magnitude as follows~\cite{Giunti:2007ry}
\begin{flalign}
	\ell_\nu = \frac{1}{n\,\sigma_\nu} \sim \frac{1}{n\,G_{\,\rm F} E_\nu M} \sim  \frac{10^{38}~{\rm cm}}
	%{(n\times1\,{\rm cm}^3)(E_\nu M/{1\,\rm GeV}^2)}.
	{n\,[{\rm cm}^{-3}]\, E_\nu\,[{\rm GeV}]\, M\,[{\rm GeV}]},
\end{flalign}
where $G_{\,\rm F}$ is the Fermi constant, $E_\nu$ is the neutrino energy, and $\sigma_\nu$ is the effective cross-sectional area for collisions of neutrinos with the target particles, which number density and mass we denote by $n$ and $M$, respectively. For usual atomic matter inside the Earth, the main target particles are nucleons with $M_n\approx1$\,GeV and $n \sim N_{\rm A}/{\rm cm}^3 \sim 10^{24}\,{\rm cm}^{-3}$. Then the Earth's ``optical density'' to the neutrino flux can be estimated as
\begin{flalign}
	\ell_\nu^{-1} \sim 10^{-9}E_\nu\,[{\rm GeV}]~{\rm km}^{-1} \sim  \frac{10^{-2}E_\nu\,[{\rm TeV}]}{D_\oplus},
\end{flalign}
where $D_\oplus\approx 12800$~km is the Earth's diameter. According to the Beer--Lambert law, this results in the flux reduction by 1\% for the neutrinos with $E_\nu\approx$~1~TeV crossing the entire Earth, while flux reduction of GeV-energy neutrinos is negligible.

For a model case with two neutrino flavors, their probability to oscillate 
depending on the distance $x$ traveled in vacuum reads
\begin{flalign}\label{eq:P}
	P_{\rm osc}(x) = \sin^22\theta \sin^2\left( \frac{\pi x}{L^{\rm osc}} \right),
\end{flalign}
where $\theta$ is the neutrino mixing angle, and
\begin{flalign}\label{eq:Losc}
	L^{\rm osc} = 2.47\, \frac{E_\nu~[{\rm GeV}]}{\Delta m^2~[{\rm eV}^2]}~\,{\rm km},
\end{flalign}
where the numerical factor comes from the relation among the chosen units and  $\Delta m^2$ is the difference between larger and smaller neutrino masses squared.

\subsection{Accounting for the matter effect}
To describe neutrinos passing through medium, the following substitution is applied to Eqs.~\eqref{eq:P} and \eqref{eq:Losc}~\cite{Wolfenstein:1977ue,Giunti:2007ry}
\begin{flalign}
	\Delta m^2 &\quad\to\quad  \Delta m^2_M = \sqrt{(\Delta m^2\cos2\theta-A_{\rm CC})^2 + (\Delta m^2\sin2\theta)^2}, \label{eq:splitting_M}\\
	\sin2\theta &\quad\to\quad  \sin2\theta_M = \frac{\Delta m^2\sin2\theta}{\Delta m^2_M},\label{eq:mixing_M}
\end{flalign}
where the matter contribution $A_{\rm CC}$ is dominated by the charged current interactions and determined through the electron neutrino potential $V_e$ as follows
\begin{flalign}\label{eq:A_CC}
    A_{\rm CC}=2V_eE_\nu = 2\sqrt{2}\,G_{\rm F}n_eE_\nu,
\end{flalign}
where the electron number density reads
\begin{flalign}
    n_e = \frac{N_{\rm A}Y_e\rho}{M_{\rm u}}
\end{flalign}
where $N_{\rm A}$ is the Avogadro constant, $M_{\rm u}$ is the molar mass constant, $\rho$ is the matter density for the given medium and $Y_e$ is the electron fraction, which we take to be equal to 0.466 for the Earth's core and 0.494 for the mantle~\cite{Lisi:1997yc}, while for 1LEM we take an approximate value of $Y_e=0.5$. 
From the above two equations the matter term can be expressed with three-digit precision as 
\begin{flalign}\label{eq:A_CC_numeric}
	A_{\rm CC} = 7.63  \times 10^{-5} \left(\frac{Y_e}{0.5} \right) \rho\,[{\rm g\, cm}^{-3}]\, E_\nu [{\rm GeV}]~ {\rm eV}^2,
\end{flalign}

\begin{figure}[tb]
	\centering
	\includegraphics[height=5cm]{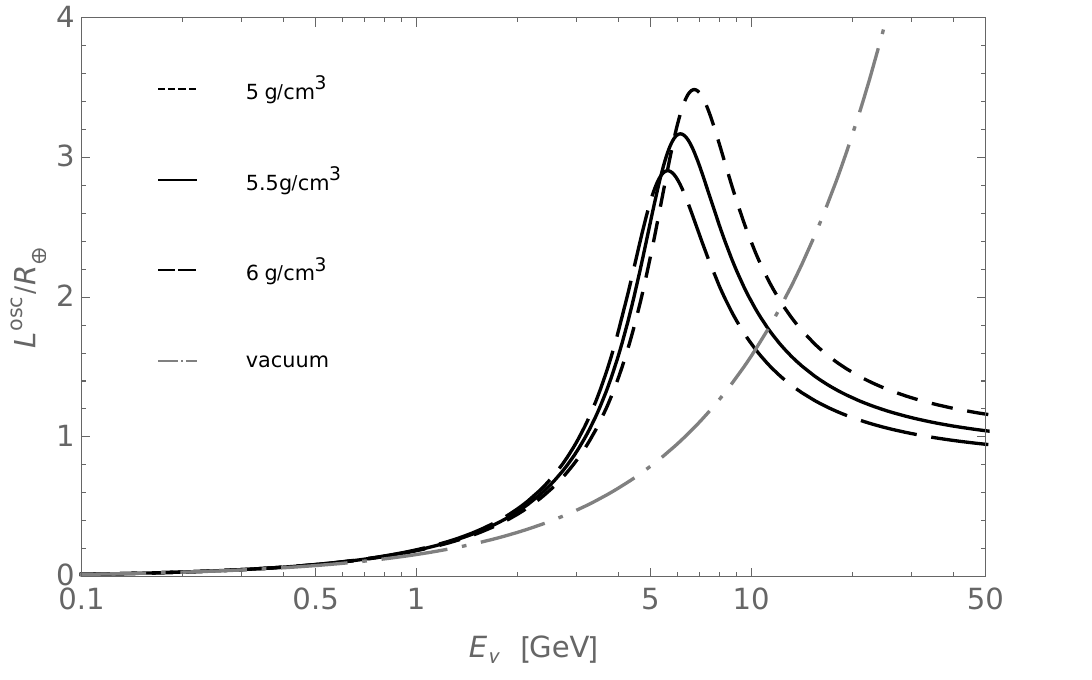}
	\caption{Two-flavor oscillation length vs energy for atmospheric neutrinos traveling through either vacuum (a simple example represented by a gray dot-dashed line) or matter with constant density $\rho$. Case of $\rho = \{5,5.5,6\}$ g\,cm$^{-3}$ is represented by \{short-dashed, solid, long-dashed\} black lines, respectively. 
	}
	\label{Fig:nu_osc_length}
\end{figure}

The oscillation length versus the neutrino energy for neutrinos traveled either in vacuum or through the matter with density $\rho=5$~g cm$^{-3}$, $\rho=5.5$~g cm$^{-3}\approx \rho_0$ and $\rho=6$~g cm$^{-3}$ is shown in Fig.~\ref{Fig:nu_osc_length}. For the atmospheric neutrino mass splitting $\Delta m^2_{31}\approx\Delta m^2_{32}$ and mixing we use the central values of $\Delta m^2_{32}=2.455\times10^{-3}$~{\rm eV}$^2$ and $\sin^2\theta_{13}=2.19\times10^{-2}$, respectively~\cite{ParticleDataGroup:2024cfk}.  
Clearly, the matter effect becomes essential for the atmospheric neutrino energies above 1~GeV and the oscillation length saturates to the value that is close to the Earth radius for $E_\nu\gg$~1~GeV. The pike at $E_\nu\sim$~6~GeV (for the discussed parameter value) takes place when $A_{\rm CC}\approx \Delta m^2\cos2\theta$ that gives $\sin2\theta_M\approx1$ and accounts for the Mikheev–Smirnov–Wolfenstein resonance enhancement of neutrino oscillation probability for certain values of the oscillation phase~\cite{Wolfenstein:1977ue,Mikheyev:1985zog,
Kelly:2021jfs}. 
We remark that three-flavor oscillation effects are suppressed for the considered parameter range~\cite{Giganti:2017fhf}.

From the above consideration this is clear that NAT method can be applied for neutrinos with energies above TeV, which flux decreases by at least 1\% due to the neutrino absorption by the Earth's matter if neutrinos come to a detector from the opposite side of the Earth. (We remark that neutrino deflections due to scatterings in the Earth core can be also essential~\cite{Borisov:1986sm}.) On the other hand, the effect of neutrino oscillations is suppressed in this case by the factor $\Delta m^2_M/\Delta m^2$. 
Contrary, for $E_\nu\sim1$~GeV the neutrino flux reduction is vanishing. However, the neutrino scattering on electrons, although being suppressed by the factor of $M_n/m_e\approx2000$, affects the fluxes of neutrinos with definite flavors by changing the oscillation picture. Hence, NOT approach alone can be safely applied in this case, which is the subject of our present consideration.

Thus NOT and NAT are effective for the neutrino energies of the order of 1~GeV and few~TeV, correspondingly. 
However, in spite of theoretical simplicity of NAT approach, 
its capabilities and accuracy are limited by available flux of the atmospheric neutrinos that drops sharply with the neutrino energy increase. In particular, this flux decreases by six orders of magnitude with $E_\nu$ growth from 1~GeV to 1~TeV~\cite{Agrawal:1995gk}. Nevertheless, one cubic kilometer working volume of the huge sub-ice detector IceCube allowed recently to obtain independent information about the Earth density profile and ``weight'' the Earth with neutrinos for the first time~\cite{Donini:2018tsg}. 
Detectors of GeV-energy neutrinos require larger number density of costly sensors and mainly for this reason have smaller working volume. However, experimental power of NOT rapidly increases with collecting more data at existing neutrino facilities, e.g.,  IceCube/DeepCore~\cite{IceCubeCollaboration:2024ssx,Chattopadhyay:2025ulr}, and developing new generation neutrino detectors. In particular, Hyper-K~\cite{Jesus-Valls:2024tgd} and DUNE~\cite{Kelly:2021jfs} are under construction, and next proposals such as KM3NeT ORCA~\cite{Maderer:2022toi} and INO/ICAL~\cite{Raikwal:2023jkf} %2405.04986
are intensively discussed.

\subsection{Neutrino oscillograms at zero order}

\begin{figure}[tb]
	\centering
	\includegraphics[height=4cm]{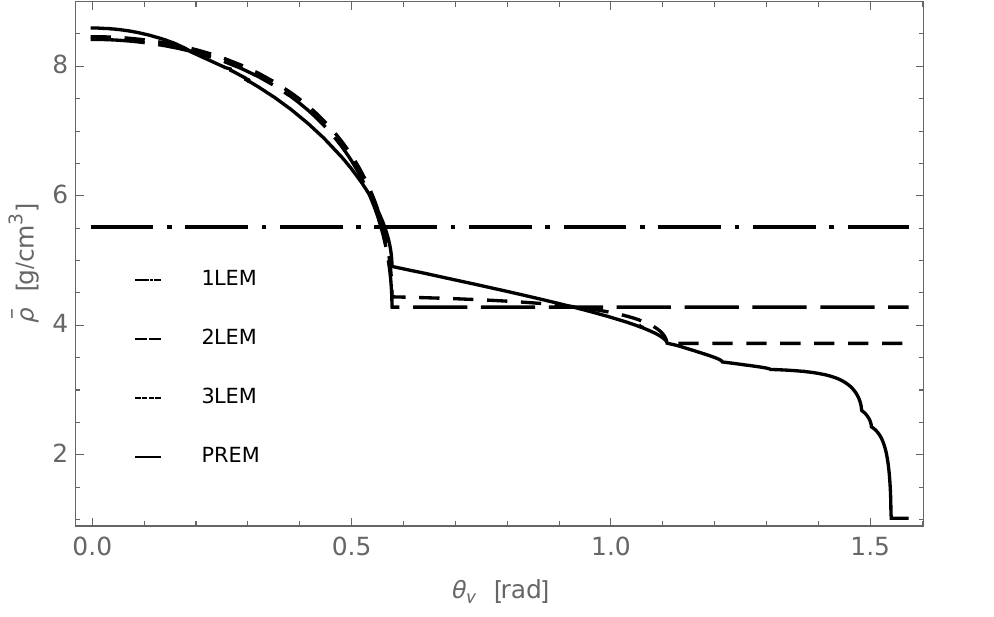}
    \includegraphics[height=4cm]{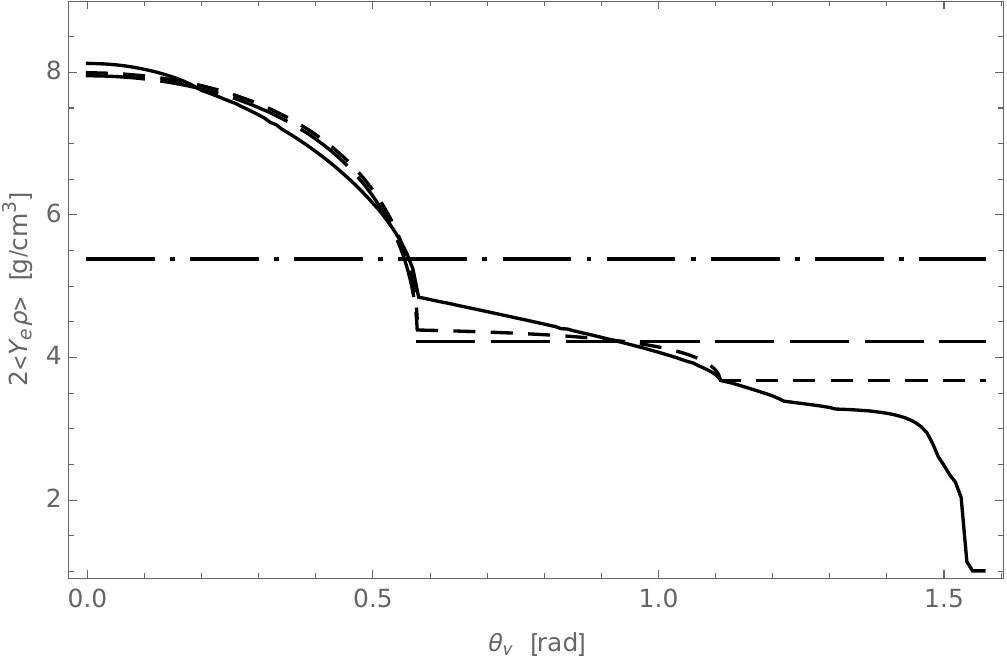}
	\caption{Average density $\bar\rho$ ({\it left}) and average product $\langle2Y_e\rho\rangle$ ({\it right}) along the neutrino path inside the Earth vs nadir angle $\theta_\nu$ for 1-, 2-, 3-layer and PREM models as represented by dot-dashed, long-dashed, short-dashed and solid lines, respectively.}
	\label{Fig:rho-theta_nu}
\end{figure}

In this section we ignore matter density variation along a given neutrino trajectory and consider neutrino passage through a medium with the constant density $\bar\rho$, which is obtained by averaging $\rho$ along the neutrino path and therefore depends on the neutrino path length $L$ inside the Earth, except for the one-layer model. 
The average density along the neutrino pass in terms of the nadir angle $\theta_\nu$ reads
\begin{flalign}\label{eq:rho_aver}
	\bar\rho(\theta_\nu) = \frac{1}{L(\theta_\nu)}\int\limits_0^{L(\theta_\nu)}\rho(x,\theta_\nu)dx,
\end{flalign}
where $L(\theta_\nu) = 2R_\oplus\cos(\theta_\nu)$, the coordinate $x$ is measured along the neutrino path, and $\theta_\nu$ is measured between the neutrino momentum and the direction to detector from the Earth's center. 
In particular, for 2LEM and 3LEM, $\bar\rho$ can be rewritten as
\begin{flalign}\label{eq:rho_aver2}
	\bar\rho = \left\{
	\begin{array}{cll}
		%\frac{(\rho_{\rm C}-\rho_{\rm M})\sqrt{R_{\rm C}^2-r_0^2} + \rho_{\rm M}\sqrt{R_\oplus^2-r_0^2}}{\sqrt{R_\oplus^2-r_0^2}}
		(\rho_{\rm C}-\rho_{\rm M})\frac{L_{\rm C}}{L} + \rho_{\rm M} & ~{\rm for}~ & r_0<R_{\rm C}, \\
		\rho_{\rm M} & ~{\rm for}~ & R_{\rm C}<r_0<R_\oplus 
	\end{array}  \right.
\end{flalign}
and
\begin{flalign}\label{eq:rho_aver3}
	\bar\rho = \left\{
	\begin{array}{lll}
		(\rho_{\rm C}-\rho_{\rm LM})\frac{L_{\rm C}}{L} + (\rho_{\rm LM}-\rho_{\rm UM})\frac{L_{\rm LM}}{L} + \rho_{\rm UM} & ~{\rm for}~ & r_0<R_{\rm C}, \\
		(\rho_{\rm LM}-\rho_{\rm UM})\frac{L_{\rm LM}}{L} + \rho_{\rm UM} & ~{\rm for}~ & R_{\rm C}<r_0<R_{\rm LM}, \\
		\rho_{\rm UM} & ~{\rm for}~ & R_{\rm LM}<r_0<R_\oplus, 
	\end{array}  \right.
\end{flalign}
respectively, where $L_{\rm C (LM)} = 2\sqrt{R_{\rm C (LM)}^2-r_0^2}$ is the distance traveled by neutrino inside the core (lower mantle), and 
$r_0 = R_\oplus\sin\theta_\nu$ is the impact parameter for the neutrino. 
The results for $\bar\rho$ and the average product $\langle\frac{Y_e}{0.5}\rho\rangle$, which is proportional to the average (`zero-order') neutrino potential $\langle V_e\rangle$, along the neutrino path in the Earth are shown in Fig.~\ref{Fig:rho-theta_nu} ({left}) and ({right}), respectively, for the discussed density schemes.

\begin{figure}[tb]
	%\centering
	\includegraphics[height=5.4cm]{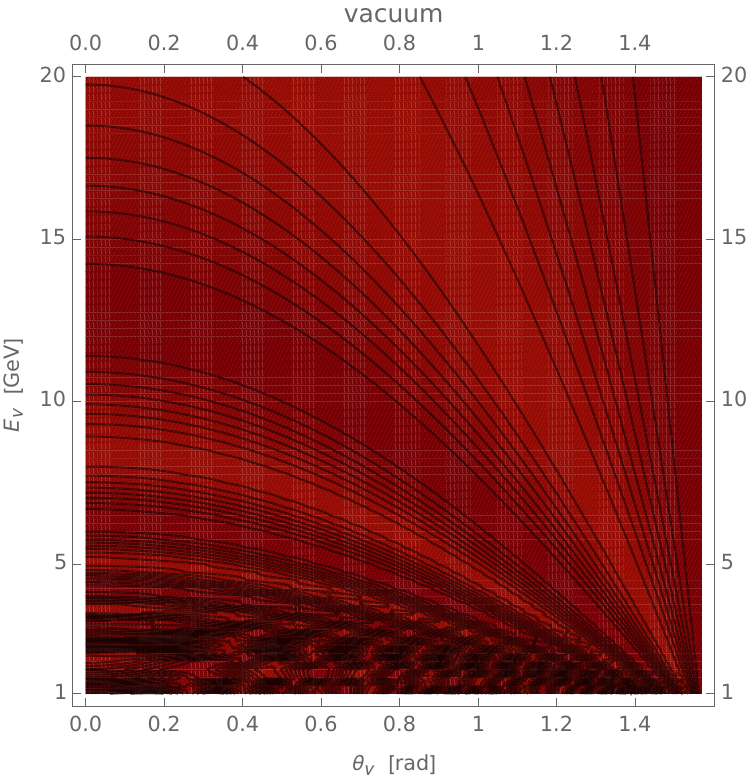}~
    \includegraphics[height=5cm]{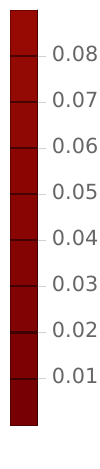}~~
	\includegraphics[height=5.4cm]{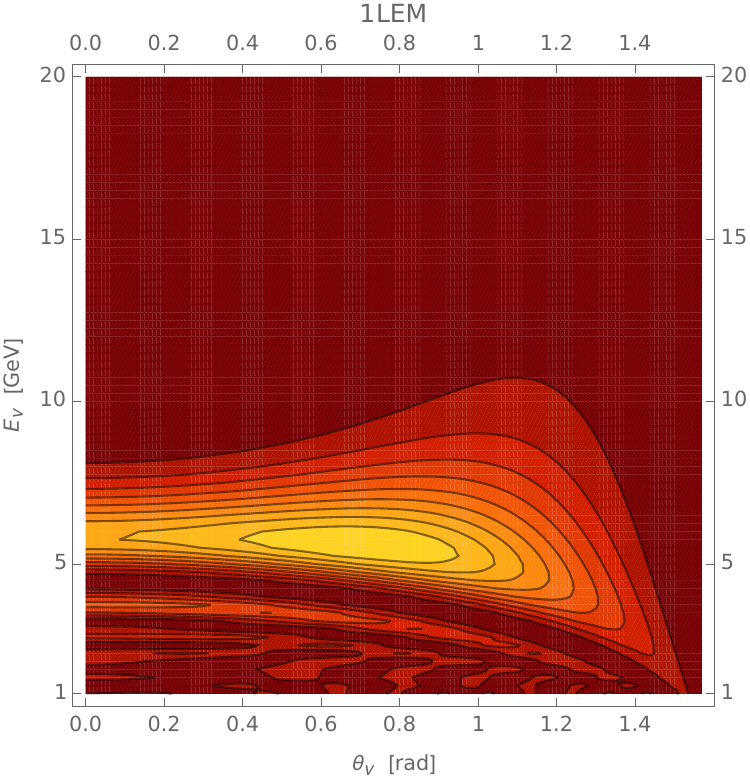}
	
	\vspace{3mm}
	
	\includegraphics[height=5.4cm]{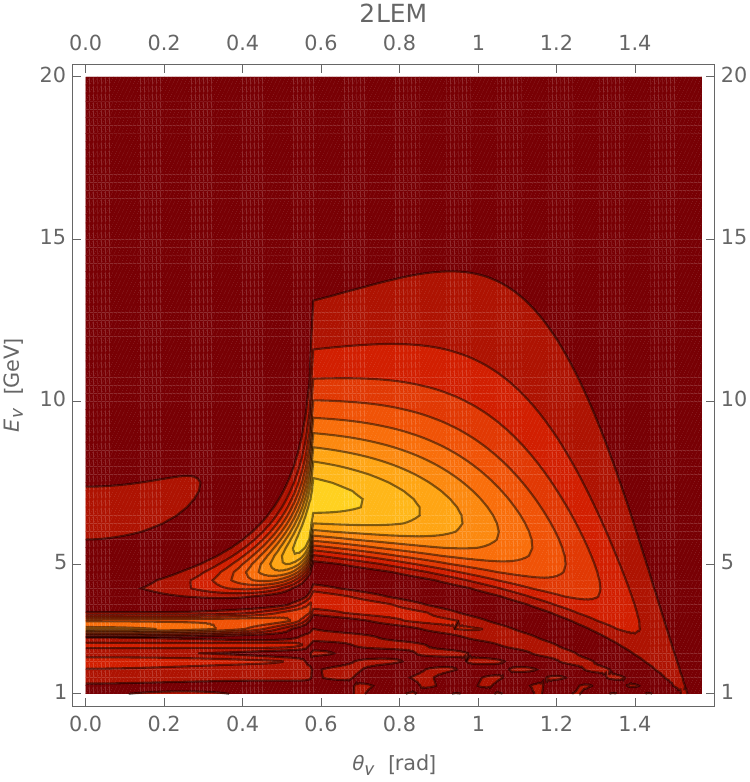}~~
	\includegraphics[height=5.4cm]{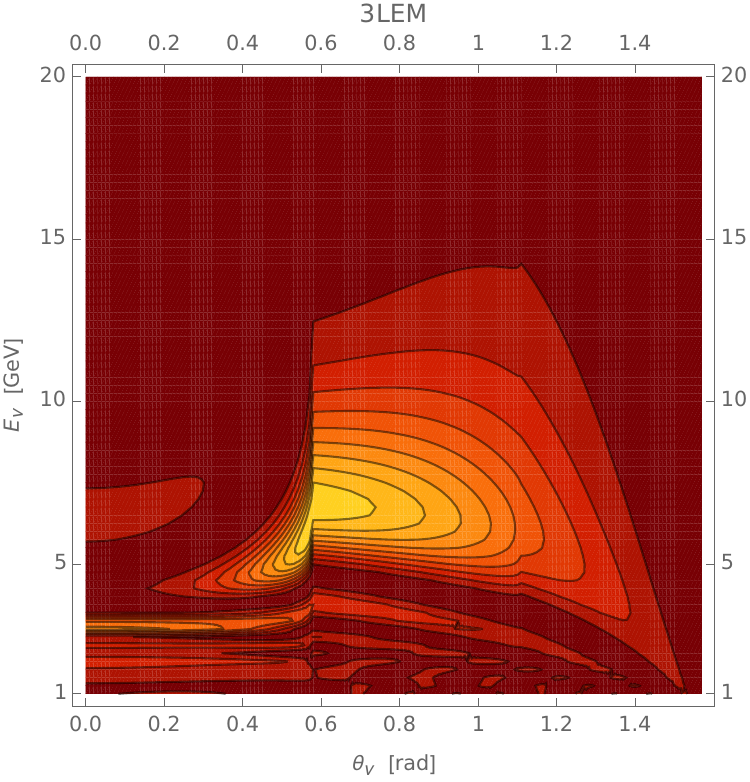}~
    \includegraphics[height=5cm]{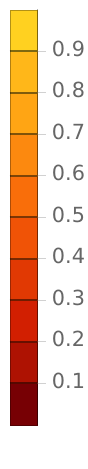}
	\caption{The neutrino oscillograms for the Earth, which density is averaged along the entire neutrino path inside the planet, in the framework of 1-layer ({\it upper right}), 2-layer ({\it lower left}) and 3-layer ({\it lower right}) Earth models. The vacuum case ({\it upper left}) is also shown for comparison.}
	\label{Fig:oscillograms0}
\end{figure}

\begin{figure}[tb]
	\centering
	\includegraphics[height=5.4cm]{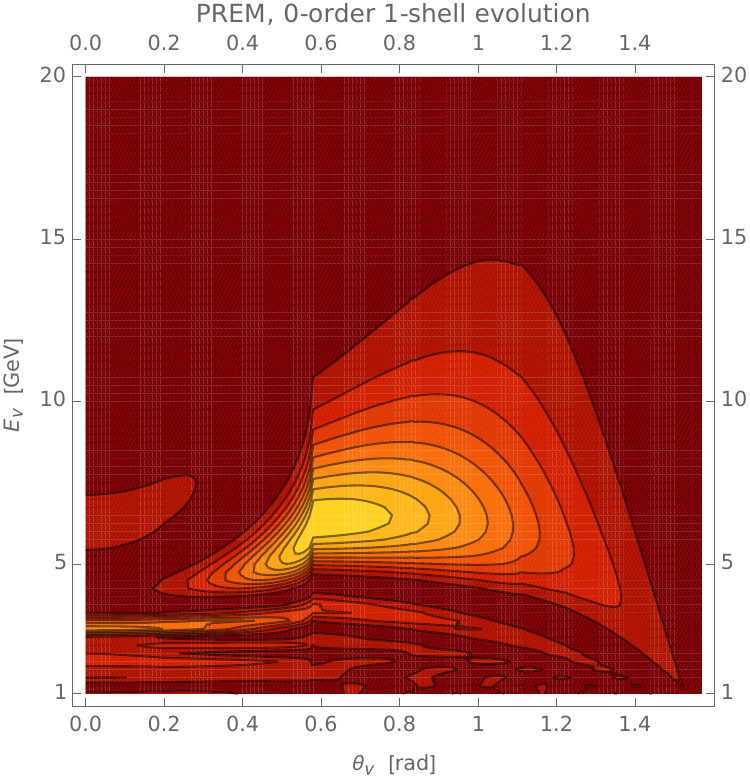}~
	\includegraphics[height=5cm]{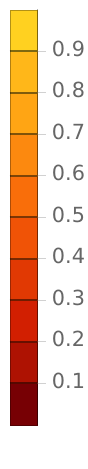}~~ 
    \includegraphics[height=5.4cm]{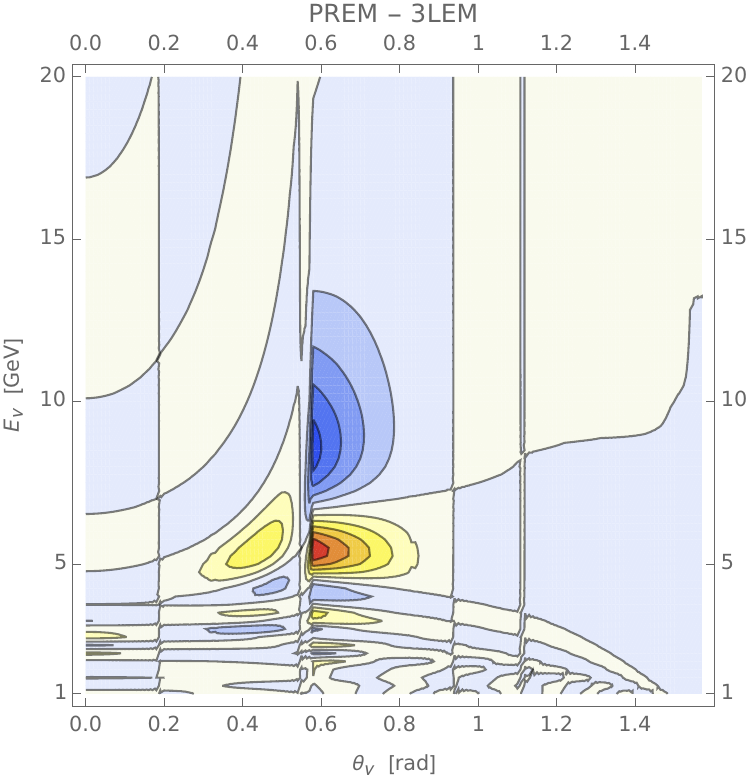}~
	\includegraphics[height=5cm]{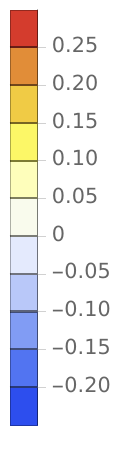}
	\caption{The zero-order oscillograms with averaging along the entire path of the neutrino (one-shell evolution) for PREM ({\it left}) and difference between PREM and 3LEM ({\it right}).}
	\label{Fig:oscillogram_PREM}
\end{figure}

This is common to visualize the oscillation probability dependence on $E_\nu$ and $\theta_\nu$ by means of 2D color maps called neutrino oscillograms. The zero-order oscillograms representing $P_{\rm osc}$ without and with matter effect (for 1-, 2- and 3-layer Earth models) are shown in Fig.~\ref{Fig:oscillograms0}. Clearly, the matter effect significantly enhances $P_{\rm osc}$ in the central and lower region of the shown parameter space. The zero-order oscillogram for PREM and its difference with the one for 3LEM are presented in Fig.~\ref{Fig:oscillogram_PREM}. In particular, it contains area with deviation in $P_{\rm osc}$ above 25\% in the lower mantle. The deviations for the trajectories crossing only the Earth's crust are relatively small in spite of significant difference in density since these trajectories are short and the accumulated matter effect is small.
In the following we take into account the effect of deviations of the neutrino potential %$V(x)$ 
from its average value within the framework of perturbation theory. 

\section{Perturbative solution for realistic density profile}\label{sec:prturbative}

The vector of neutrino flavor eigenstates $\nu_f=(\nu_e,\nu_\mu,\nu_\tau)^T$ is related to the vector of neutrino mass eigenstates $\nu_m=(\nu_1,\nu_2,\nu_3)^T$ by the unitary rotation
\begin{flalign}
    \nu_f = U\nu_m,
\end{flalign} 
where the neutrino mixing matrix $U$ contains three angles and at least one $CP$-violating phase. 
The evolution of the neutrino vector of state can be described by the Schr\"odinger-like equation, which in the Planck units reads
\begin{eqnarray}
    i\frac{d\nu_f}{dx} = %&=& H%\mathcal{H}\,%
    \left( \frac{UM^2U^\dag}{2E_\nu} + \hat V \right) 
    \nu_f,\\
    %\mathcal{H} &=& \frac{UM^2U^\dag}{2E_\nu} + \mathcal{V}
\end{eqnarray}
where $M^2={\rm diag}(0,\Delta m_{21}^2,m_{31}^2)$ is the diagonal matrix of
neutrino mass squared differences $\Delta m_{ij}^2=m_i^2-m_j^2$, and $\hat V = {\rm diag}(V_e,0,0)$ is the matrix of matter-induced neutrino potentials.

Taking into account strongly hierarchical neutrino masses~\cite{ParticleDataGroup:2024cfk}, i.e. 
\begin{flalign}
    \Delta m^2\equiv|\Delta m_{13}^2|\approx|\Delta m_{23}^2|\gg|\Delta m_{12}^2|,
\end{flalign}
we ignore the 1-2 neutrino mixing and using the neutrino propagation basis~\cite{Akhmedov:1998xq,Akhmedov:2006hb} ($\nu_\mu$ and $\nu_\tau$ propagate equally since their matter potentials are the same) rewrite the evolution equation in a two-neutrino mixing scheme
\begin{eqnarray}\label{eq:2-nu}
    i\frac{d}{dx} \left( \begin{array}{c}
			\nu_{e} \\
			\nu_{\alpha}
		\end{array} \right) = H
	\left( \begin{array}{c}
		\nu_{e} \\
		\nu_{\alpha}
		\end{array} \right)
\end{eqnarray}
where $\nu_\alpha=\nu_\mu\sin\theta_{23}+\nu_\tau\cos\theta_{23}$%(omitting the $CP$-phase that does not affect the oscillations)
, and the effective Hamiltonian %in the flavor basis 
%in the basis where ${H}_{11}=-{H}_{22}$ reads
can be written as
\begin{flalign}\label{eq:H}
    %\mathcal{\hat H} \equiv 
    {H} = \frac{1}{4E} 
    \left( \begin{array}{cc}
        -\Delta m^2\cos2\theta + A_{\rm CC} & \Delta m^2\sin2\theta \\
        \Delta m^2\sin2\theta  & \Delta m^2\cos2\theta - A_{\rm CC}
    \end{array} \right),
\end{flalign}
where the part proportional to unit matrix is omitted since it does not affect the oscillations. 
The matrix ${H}$ depends on the only mixing angle $\theta\equiv\theta_{13}$, and can be diagonalized by the orthogonal transformation
\begin{flalign}\label{eq:H_diag}
    U_M^T {H} U_M = %\mathcal{H}_M \equiv 
	\frac{1}{4E}\,{\rm diag}(-\Delta m_M^2,\Delta m_M^2),
\end{flalign}
where
\begin{flalign}\label{eq:U_M}
    U_M = \left( \begin{array}{cc}
		\cos\theta_M & \sin\theta_M \\
		-\sin\theta_M & \cos\theta_M
	\end{array} \right)
\end{flalign}
is the effective mixing matrix in matter, and $\Delta m_M^2$ and $\theta_M$ are defined in Eqs.~\eqref{eq:splitting_M} and \eqref{eq:mixing_M}. Mathematically, the considered problem with only two effective generations, when one neutrino is ``decoupled'' from the other two, is similar to the two-level system~\cite{Supanitsky:2008eq,Blanes:2008xlr}.

In the case of medium with relatively weakly varying density the following perturbative approach can be used.

\subsection{Evolution matrix. Perturbative solution}

Evolution of the neutrino state over a finite distance from 0 to $x$ 
can be described as
\begin{eqnarray}\label{eq:evolution}
    \left( \begin{array}{c}
			\nu_{e}(x) \\
			\nu_{\alpha}(x)
		\end{array} \right) = S(x)
	\left( \begin{array}{c}
		\nu_{e}(0) \\
		\nu_{\alpha}(0)
		\end{array} \right),
\end{eqnarray}
using the evolution matrix $S(x)\equiv S(x, 0)$, which satisfies the same evolution equation as the state vector
\begin{eqnarray}\label{eq:evolution_eq}
	i\frac{dS(x)}{dx} = H(x)S(x).
\end{eqnarray}
The probability of $\nu_e\leftrightarrow\nu_\alpha$ oscillations is given by $P_{e\alpha}(x)=|S_{e\alpha}(x)|^2$. Returning to the flavor basis, we can approximate the matrix of oscillation amplitudes among the three flavor neutrino states by the product~\cite{Akhmedov:1998xq,Akhmedov:2006hb}
\begin{flalign}
    \left( \begin{array}{ccc}
		1 & 0 & 0 \\
        0 & \cos\theta_{23} & \sin\theta_{23} \\
		0 & -\sin\theta_{23} & \cos\theta_{23}
	\end{array} \right)
    \left( \begin{array}{ccc}
		S_{ee} & 0 & S_{e\alpha} \\
        0 & 1 & 0 \\
		S_{\alpha e} & 0 & S_{\alpha\alpha}
	\end{array} \right)
    \left( \begin{array}{ccc}
		1 & 0 & 0 \\
        0 & \cos\theta_{23} & -\sin\theta_{23} \\
		0 & \sin\theta_{23} & \cos\theta_{23}
	\end{array} \right),
\end{flalign}
which results in the observable oscillation probabilities, in particular,  $P(\nu_e\leftrightarrow\nu_\mu)=P_{e\alpha}\sin^2\theta_{23}$ and $P(\nu_e\leftrightarrow\nu_\tau)=P_{e\alpha}\cos^2\theta_{23}$.

In the perturbative approach the matter-induced neutrino potential, $V\equiv V_e$ for short, can be written as
\begin{flalign}
    V(x) = \bar V + \Delta V(x),
\end{flalign}
where
\begin{flalign}\label{eq:barV}
    \bar V \equiv \langle V(x)\rangle = \frac{1}{L}\int_0^L V(x) dx
\end{flalign}
is the average value of the potential along the entire neutrino path in the Earth and $\Delta V(x)$ is a relatively small perturbation. %Correspondingly, 
Then the Hamiltonian
\begin{flalign}\label{eq:H(x)}
    H(x) = \bar H + \Delta H(x)
\end{flalign} 
can be expressed using the Pauli matrices $\sigma_1$ and $\sigma_3$:
\begin{flalign}\label{eq:H(x)_2}
    \bar H = \bar\omega( \sigma_1 \sin2\bar\theta - \sigma_3 \cos2\bar\theta ), 	\qquad\qquad  	\Delta H(x) = \frac{\Delta V(x)}{2} \sigma_3,
\end{flalign}
where the constant term $\bar H\equiv H(\bar V)$ is obtained from Eqs.~\eqref{eq:H_diag} and \eqref{eq:U_M}, $\bar\theta\equiv\theta_M(\bar V)$ is the mixing angle in matter,
\begin{flalign}
    \bar\omega \equiv \omega(\bar V) = \frac{\Delta m_M^2(\bar V)}{4E_\nu} = \frac{\pi}{\bar L^{\rm osc}}
\end{flalign}
is the half-frequency in matter with $\bar L^{\rm osc}\equiv L^{\rm osc}(\Delta m_M^2(\bar V))$, and $\Delta H(x)$ follows from Eqs.~\eqref{eq:H} and \eqref{eq:A_CC}. For the case of constant density, %$H(x) = \bar H$, In the limit of 
i.\,e. vanishing $\Delta V$, the Hamiltonian in Eq.~\eqref{eq:H(x)} takes to form of $H(\bar V)$ in Eq.~\eqref{eq:H}, 
and solution of Eq.~\eqref{eq:evolution_eq} can be 
written as the matrix exponential of Pauli matrices, 
\begin{flalign}\label{eq:Sbar}
    \bar S(x) = e^{-i\phi(x)\hat n\cdot{\vec\sigma}} = c_{\phi(x)}\mathbb{I} - i s_{\phi(x)} \hat n\cdot{\vec\sigma},
\end{flalign}
where we denote $c_\phi\equiv\cos\phi$, $s_\phi\equiv\sin\phi$ for short, $\phi(x)=\bar\omega x$ is the half-phase of oscillations in matter, and the unit vector is given by
\begin{flalign}
    \hat n = (s_{2\bar\theta},0,-c_{2\bar\theta}).
\end{flalign}
Obviously, $|\bar S_{e\alpha}(x)|^2$ equals to $P_{\rm osc}(x)$ in Eq.~\eqref{eq:P}.

Seeking the solution of Eq.~(\ref{eq:evolution_eq}) in the form~\cite{Akhmedov:2006hb}
\begin{flalign}
    S(x) = \bar S(x) + \Delta S(x),
\end{flalign}
where  
\begin{flalign}\label{eq:DeltaS}
    \Delta S(x) = -i\bar S(x)K(x),
\end{flalign}
we obtain the following equation on $K(x)$
\begin{eqnarray}\label{eq:evolution_eq_K}
	\frac{dK(x)}{dx} = A(x) (\mathbb{I} - iK(x)), 
\end{eqnarray}
where $\mathbb{I}$ is the unit $2\times2$ matrix, and 
\begin{flalign}\label{eq:A}
    A(x) = \bar S^\dag\Delta H(x)\bar S.
\end{flalign}
Assuming $|K(x)_{ab}|\ll 1$, we look for an iterative solution of
Eq.~\eqref{eq:evolution_eq_K} in the form
\begin{flalign}\label{eq:K}
    K(x) = K_1(x) + K_2(x) + \dots, %+ K_n(x),
\end{flalign}
where $|(K_{n+1})_{ab}|<|(K_{n})_{ab}|$ and the terms in the right-hand side satisfy the following equations
\begin{flalign}
    \frac{dK_1(x)}{dx} &= A(x), \label{eq:K1}\\
	\frac{dK_j(x)}{dx} &= -iA(x)K_{j-1}(x), \quad  j=2,3,\dots \label{eq:Kj}
\end{flalign}
Solution of this system, which derivation is discussed in the Appendix~\ref{sec:A1}, reads
\begin{flalign}\label{eq:K_solution}
	K(x) = -i \left( \mathbb{I} - \mathcal{T}e^{-i\int\limits_0^{x} d\xi A(\xi)} \right).
\end{flalign}
The resulting evolution matrix
\begin{flalign}%\label{eq:Sbar}
    S(x) = e^{-i\phi(x)\hat n\cdot{\vec\sigma}} \mathcal{T}e^{-i\int\limits_0^{x} d\xi A(\xi)}
\end{flalign}
is unitary due to hermeticity of $\Delta H$.

\subsection{First order}\label{sec:order1}

By integrating the equation
\begin{flalign}\label{eq:K1(x)}
    \frac{dK_1(x)}{dx} = \frac{\Delta V(x)}{2}\,\bar S^\dag\sigma_3\bar S
\end{flalign}
along the neutrino path for the unit vector $\hat n$ with zero second component we obtain
\begin{flalign}
    K_1(L) \equiv \int_0^L K_1(x)dx = \frac{n_1}{2}\, e^{i\bar\omega\,\hat n\cdot{\vec\sigma}L%\frac{L}{2}
    } \left( -[\hat n\times\vec\sigma]_2\, {\rm Re}(I)%\Delta I 
    + \sigma_2\, {\rm Im}(I)%\Delta J 
    \right),
\end{flalign}
where 
\begin{flalign}\label{eq:I}
    I%\Delta I 
    \equiv \int_{-L/2}^{L/2} \Delta V\left(x+\frac{L}{2}\right)e^{i2\bar\omega x} dx
    = e^{-i\bar\omega L} \int_{0}^{L} \Delta V(x)e^{i2\bar\omega x} dx
\end{flalign}
and the details of derivation are given in the Appendix~\ref{sec:A2}. We remark that ${\rm Re}(I)$ and ${\rm Im}(I)$ represent, correspondingly, effects of symmetric and antisymmetric components of $\Delta V$ with respect to the midpoint of the neutrino trajectory. 
Then the first-order perturbation of the evolution matrix reads
\begin{flalign}\label{eq:DeltaS1}
    \Delta S_1(L) \equiv -i\bar S(L)K_1(L) = i\frac{n_1}{2} \left( [\hat n\times\vec\sigma]_2 \,{\rm Re}(I) - \sigma_2 \,{\rm Im}(I) \right),
\end{flalign}
and, introducing a unit vector
\begin{flalign}\label{eq:k1vec}
    \hat{k}_1= \frac{1}{|I|}\left(-n_3{\rm Re}(I),{\rm Im}(I),n_1{\rm Re}(I)%\Re(I)
    \right)
\end{flalign}
that is orthogonal to $\hat n$, we obtain 
\begin{flalign}\label{eq:S_first-order}
    S_1(L) \equiv \bar S+\Delta S_1(L) = e^{-i\bar{\omega}L\hat{n}\cdot\vec{\sigma}}-i\varepsilon_1(L)\,\hat{k}_1\cdot\vec{\sigma},
\end{flalign}
where $%\varphi
\varepsilon_1=n_1|I|/2$, and the following equality ensures the unitarity of the evolution matrix at the first order in $\Delta S$ in the considered approximation (this was also discussed in Ref.~\cite{Akhmedov:2006hb})
\begin{flalign}\label{eq:unitarity_deviation1}
    |S_1(L)|^2-\mathbb{I} = \varepsilon_1(L)^2\,\mathbb{I} = \mathcal{O}\left(\xi^2\right),
\end{flalign}
where $\xi=|\Delta V/\bar V|$ and $\mathcal{O}\left(\xi^2\right)$ denotes the terms that are determined by the 2nd power of $\xi$.  
Although, finite-order iterations of the evolution matrix do not have to preserve unitarity exactly, Eq.~\eqref{eq:S_first-order} for small $\varepsilon_1$ can be approximated by the following unitary expression
\begin{flalign}\label{eq:S}
    S_1(L)\approx c_{\varepsilon_1} e^{-i\bar{\omega}L\hat{n}\cdot\vec{\sigma}}-is_{\varepsilon_1}\hat{k}_1\cdot\vec{\sigma} = c_{\varepsilon_1}c_{\bar{\omega}L}\mathbb{I}-i\left(c_{\varepsilon_1} s_{\bar{\omega}L}\hat{n}+s_{\varepsilon_1}\hat{k}_1\right)\cdot\vec{\sigma}.
\end{flalign}

\subsection{Second order}

To the second order we have
\begin{flalign}
    \Delta S(L) = -i\bar S(L)[K_1(L)+K_2(L)],
\end{flalign}
where $K_1(L)$ was obtained in the previous section %~\ref{sec:order1}, 
and
\begin{flalign}
    K_2(L) &=  -i \int_0^L dx_2 A(x_2) \int_0^{x_2} dx_1 A (x_1) \nonumber\\
    &= -i \int_{-L/2}^{L/2} dy A\left( y+\frac{L}{2} \right) \int_0^{y+\frac{L}{2}} dx A (x).
\end{flalign}
Using Eq.~\eqref{eq.4.5} and commutativity of $\bar S(x)$ with $\hat n\cdot{\vec\sigma}$, $K_2(L)$ can be rewritten as
\begin{eqnarray}\label{eq:K2}
	K_2(L) =  -\frac{i}{4}\bar S^\dag(L) \int\limits_{-L/2}^{L/2} dy \Delta V\left( y+\frac{L}{2} \right) \int\limits_0^{y+\frac{L}{2}} dx \Delta V(x) 
	F(x,y),
\end{eqnarray}
where the matrix function $F$ is given by%can be written as
\begin{flalign}
    F(x,y) =&  
	\left( n_1 
	\left( -c_{2\bar\omega y} [\hat n\times\vec\sigma]_2 + s_{2\bar\omega y} \sigma_2 
	\right) 
	+ n_3 \bar S(L) \hat n\cdot{\vec\sigma}
	\right) \nonumber\\
	\cdot&\left( n_1 
	\left( -c_{2\bar\omega x} [\hat n\times\vec\sigma]_2 + s_{2\bar\omega x} \sigma_2 
	\right) 
	+ n_3 \hat n\cdot{\vec\sigma}
	\right).
\end{flalign} 

To shorten subsequent expressions, we introduce the double integral
\begin{flalign}\label{eq:J}
    J\left(f
    \right) &\equiv \int\limits_{-L/2}^{L/2} dy\Delta V\left( y+\frac{L}{2} \right) \int\limits_0^{y+\frac{L}{2}} dx\Delta V(x) f\left( x,y \right) 
    \nonumber\\
    & = \int\limits_0^L dt\Delta V(t) \int\limits_0^{t} dx\Delta V(x) f\left( x,t-\frac{L}{2} \right)
\end{flalign}
and notice that $J(1)=-J(1)=0$ for any form of the potential due to $\langle \Delta V\rangle=0$ and equivalence of triangular integration regions $\{(x,t)\,|\,x\in(t,L),\,t\in(0,L)\}$ and $\{(x,t)\,|\,x\in(0,L),\,t\in(0,x)\}$. 
Then, using the expression for $\bar S$ in Eq.~(\ref{eq:Sbar}), the pairwise anticommutativity among $\hat n\cdot{\vec\sigma}$, $\sigma_2$ and $[\hat n\times\vec\sigma]_2$, the products of
\begin{flalign}
    (\hat n\cdot{\vec\sigma}) \,\sigma_2 = - i\, [\hat n\times\vec\sigma]_2, \quad (\hat n\cdot{\vec\sigma})\, [\hat n\times\vec\sigma]_2 = i\sigma_2, \quad  [\hat n\times\vec\sigma]_2\, \sigma_2 = i (\hat n\cdot{\vec\sigma}),
\end{flalign}
and ignoring the constant terms %(proportional to unit matrix) 
that disappear when substituted into Eq.~\eqref{eq:K2}, we rewrite the matrix function as
\begin{flalign}
    F(x,y) &= n_1^2 \left( c_{2\bar\omega (y-x)}\mathbb{I} + i s_{2\bar\omega (y-x)} \hat n\cdot{\vec\sigma} \right) 
	+ i\, n_1n_3  \left( c_{2\bar\omega y}\sigma_2 + i s_{2\bar\omega y} [\hat n\times\vec\sigma]_2 \right) \nonumber\\
	&- i\, n_1n_3 %\bar S(L) 
	\left( c_{\bar\omega L}\mathbb{I} - i s_{\bar\omega L} \hat n\cdot{\vec\sigma} \right)
	\left( c_{2\bar\omega x}\sigma_2 + i s_{2\bar\omega x} [\hat n\times\vec\sigma]_2 \right),
\end{flalign}
which results in
\begin{flalign}
	\Delta S_2(L) &\equiv -i\bar S(L)K_2(L) =  -\frac{1}{4} \left\{ n_1^2 \left[ J(c_{2\bar\omega (x-y)}) \mathbb{I} - iJ(s_{2\bar\omega (x-y)})\hat n\cdot{\vec\sigma} \right] \right. \nonumber\\
	&+ \left. i\,n_1n_3 \left[ J(c_{2\bar\omega y} - c_{2\bar\omega (x-L/2)}) \sigma_2 + J(s_{2\bar\omega y} - s_{2\bar\omega (x-L/2)})[\hat n\times\vec\sigma]_2 \right] \right\}.
\end{flalign}
Using the equality $J\left(e^{i2\bar\omega y}\right) = -J\left(e^{i2\bar\omega (x-L/2)}\right)$, we come up with
\begin{flalign}
	\Delta S_2(L) %\equiv -i\bar S(L)K_2(L) 
    &=  -\frac{n_1^2}{4} \left[ %\mathcal{I}
    J(c_{2\bar\omega (x-y)}) \mathbb{I} - i%\,\mathcal{I}
    J(s_{2\bar\omega (x-y)})
    \hat n\cdot{\vec\sigma} \right] \nonumber\\
    &- i\frac{n_1n_3}{2} \left[ %\mathcal{I}
    J(c_{2\bar\omega y}) \sigma_2 + %\mathcal{I}
    J(s_{2\bar\omega y})[\hat n\times\vec\sigma]_2 \right].
\end{flalign}
Thus, denoting $J\equiv J\left(e^{i2\bar\omega y}\right)$ for short and defining, similarly to $\hat k_1$ in Eq.~\eqref{eq:k1vec}, a unit vector %$\hat{k}_2 = \vec{K}/|\vec{K}|$, where 
\begin{flalign}
    \hat{k}_2 = \frac{1}{\left|J%\left(e^{i2\bar\omega y}\right)
    \right|} \left(n_3%J(s_{2\bar\omega y})
    {\rm Im}J, %J(c_{2\bar\omega y})
    {\rm Re}J, -n_1 %J(s_{2\bar\omega y}
    {\rm Im}J) \right)
\end{flalign}
orthogonal to $\hat n$, we obtain the second-order iteration of the evolution matrix
\begin{flalign}\label{eq:S_second-order}
    S_2(L) &\equiv \bar S+\Delta S_1(L)+\Delta S_2(L) 
   %= e^{-i\bar\omega L\hat{n}\cdot\vec{\sigma}}
    = \left(c_{\bar\omega L} - \frac{n_1^2}{4}J(c_{2\bar\omega(x-y)})\right)\mathbb{I} \nonumber\\
    &- i \left[ \left(s_{\bar\omega L} - \frac{n_1^2}{4}J(s_{2\bar\omega(x-y)})\right)\hat n + \frac{n_1}{2}\left( |I|\hat{k}_1 + n_3|J%\left(e^{i2\bar\omega y}\right)
    |\hat{k}_2 \right) \right]\cdot{\vec\sigma}.
\end{flalign}
Taking into account that the integrands of $I$ and $J$ are first- and second-order in $\Delta V$, respectively, and 
\begin{flalign}\label{eq:unitarity_deviation2}
    |S_2|^2-\mathbb{I} 
    &=  \frac{n_1^2n_3}{2} \left[ {\rm Im}(I)J(c_{2\bar\omega(x-y)}) - {\rm Re}(I)J(s_{2\bar\omega(x-y)}) \right] \nonumber\\
    &+ \mathcal{O}\left( \xi%\left(\frac{\Delta V}{\bar V}\right)
    ^4\right) = \mathcal{O}\left( \xi%\left(\frac{\Delta V}{\bar V}\right)
    ^3\right),
\end{flalign}
we find that deviation from unitarity is reduced %from the second to the third order in $\Delta V$ 
in this case with respect to the first-order result in Eq.~\eqref{eq:unitarity_deviation1}. Moreover, for specific parameter values the expression in the square brackets in Eq.~\eqref{eq:unitarity_deviation2} can vanish [to check]. 
We remark that the term $\varepsilon_1(L)^2\,\mathbb{I}$ cancels out in Eq.~\eqref{eq:unitarity_deviation2} due to the equality
\begin{flalign}\label{eq:I-J_relation}
    |I|^2 = 2 \left[ c_{\bar\omega L}J(c_{2\bar\omega(x-y)}) +  s_{\bar\omega L}J(s_{2\bar\omega(x-y)})\right],
\end{flalign}
which can be proven using the change of integration limits discussed after Eq.~\eqref{eq:J}. 

The expression for the evolution matrix in Eq.~\eqref{eq:S_second-order} can be rewritten as 
%[to check positivity of the expressions under square roots]
\begin{flalign}\label{eq:S_second-order_v2}
    S_2 
    = \sqrt{1-\varepsilon_1^2+\varepsilon_2^2}\, (c^\prime\mathbb{I} - i s^\prime \hat n \cdot{\vec\sigma}) - i \sqrt{\varepsilon_1^2+\varepsilon^{\prime2}} \, \hat{k}^\prime\cdot{\vec\sigma},
\end{flalign}
where the functions $\varepsilon_2 = n_1^2|J|/4$ and
\begin{flalign}
     \varepsilon^{\prime2} = \frac{n_1^2n_3}{4} \left[ n_3|J|^2 + 2{\rm Im}(IJ^*)
     %{\rm Im}(I){\rm Re}(J) - {\rm Re}(I){\rm Im}(J) 
     \right]
\end{flalign}
come from the second-order contribution (the parameter $\epsilon'$ has either real or purely imaginary values), the functions
\begin{flalign}
     c^\prime = \frac{c_{\bar\omega L} - \frac{n_1^2}{4}J(c_{2\bar\omega(x-y)})}{\sqrt{1-\varepsilon_1^2+\varepsilon_2^2}}, \qquad
     s^\prime = \frac{s_{\bar\omega L} - \frac{n_1^2}{4}J(s_{2\bar\omega(x-y)})}{\sqrt{1-\varepsilon_1^2+\varepsilon_2^2}}
\end{flalign}
represent modification of $c_{\bar\omega L}$ and $s_{\bar\omega L}$ in the second-order iteration and satisfy the equation $c^{\prime2} + s^{\prime2} = 1$ due to Eq.~\eqref{eq:I-J_relation}, and the new unit vector $\hat{k}^\prime$ is defined by
\begin{flalign}
    A\, \hat{k}^\prime &= |I| \,\hat{k}_1 + n_3 |J| \,\hat{k}_2, \nonumber\\ 
    A &= \sqrt{|I|^2 + 2n_3{\rm Im}(IJ^*) + n_3^2|J|^2}.
\end{flalign}
Now, unitarization of Eq.~\eqref{eq:S_second-order_v2} reads
\begin{flalign}\label{eq:S_2}
    S_2(L) 
    \approx \sqrt{c_{\varepsilon^\prime}^2 c_{\varepsilon_2}^2-\varepsilon_1^2+s_{\varepsilon_2}^2}\, (c^\prime\mathbb{I} - i s^\prime \hat n \cdot{\vec\sigma}) - i \sqrt{\varepsilon_1^2+s_{\varepsilon^\prime}^2c_{\varepsilon_2}^2} \, \hat{k}^\prime\cdot{\vec\sigma}.
\end{flalign}

\section{Application to realistic Earth density distribution}\label{sec:evolution}
\subsection{One-shell evolution}

\begin{figure}[tb]
	\centering
	\includegraphics[height=5.4cm]{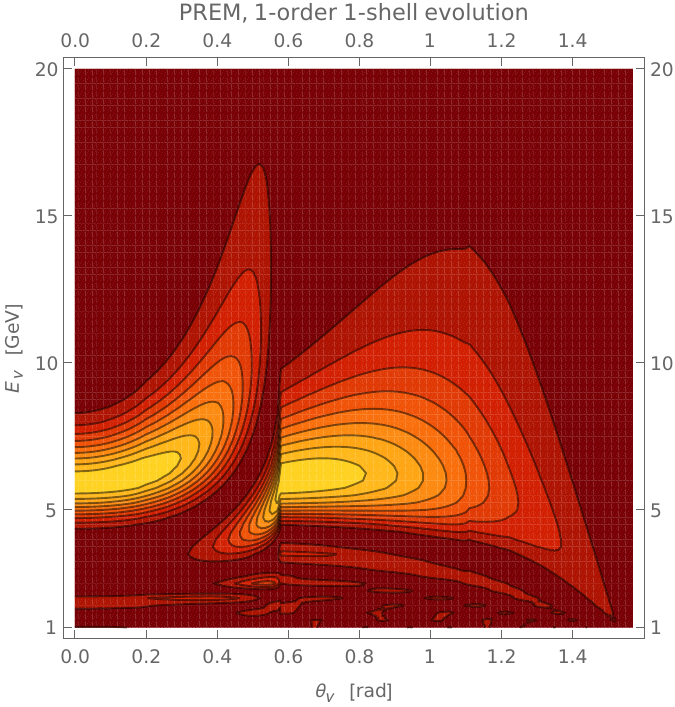}~
	\includegraphics[height=5cm]{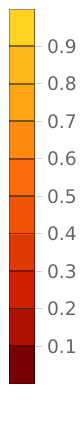}~~
    \includegraphics[height=5.4cm]{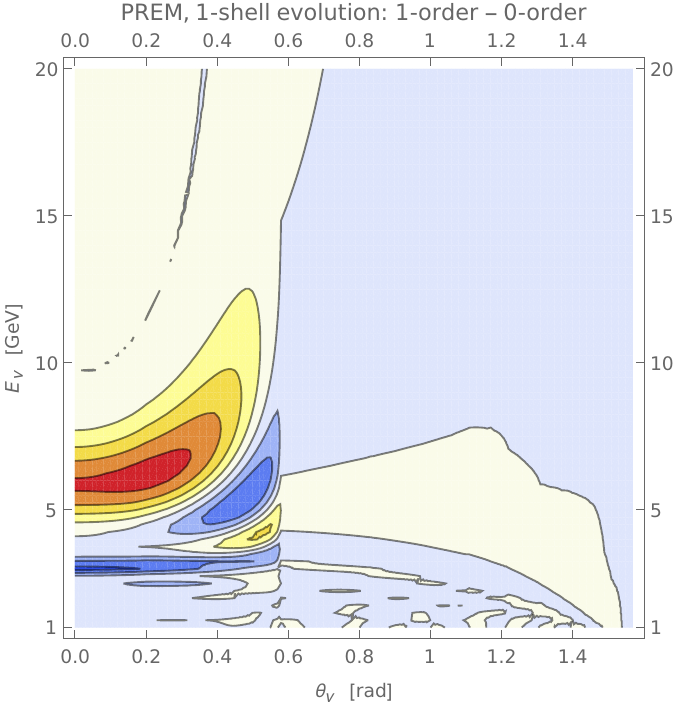}~
	\includegraphics[height=5cm]{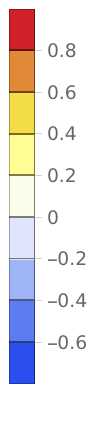}
	\caption{The oscillograms for PREM for neutrinos evolved in a single shell: the first-order perturbative result ({\it left}) and difference between the first-order and zero-order iterations ({\it right}).
    }
	\label{Fig:oscillogram_PREM_1-order_1-shell}
\end{figure}

Considering the entire Earth as a single shell, we describe the neutrino state evolution on its path $L$ by a single matrix $S(L)$ in Eq.~\eqref{eq:evolution}. The results for the oscillation probability in the zero-order approximation, i.\,e. $P_{ea}(L)=|\bar S_{ea}(L)|^2$, are shown in Figs.~\ref{Fig:oscillograms0} and \ref{Fig:oscillogram_PREM} for several models of the Earth's density distribution. 
The unitarized first-order perturbative result, $P_{ea}(L)=|S_{1}(L)_{ea}|^2$, for PREM and its comparison to the zero-order one is shown in Fig.~\ref{Fig:oscillogram_PREM_1-order_1-shell}. The first-order result significantly differs from the zero-order one for the core-crossing neutrino trajectories, for which the difference can even exceed 80\%. In fact, this is related to the breaking down of the condition $|K_{ab}|\ll1$ 
due to the large density jump between mantle and core, in other words the perturbation theory does not work in this case.

\begin{figure}[tb]
	\centering
	\includegraphics[height=5.2cm]{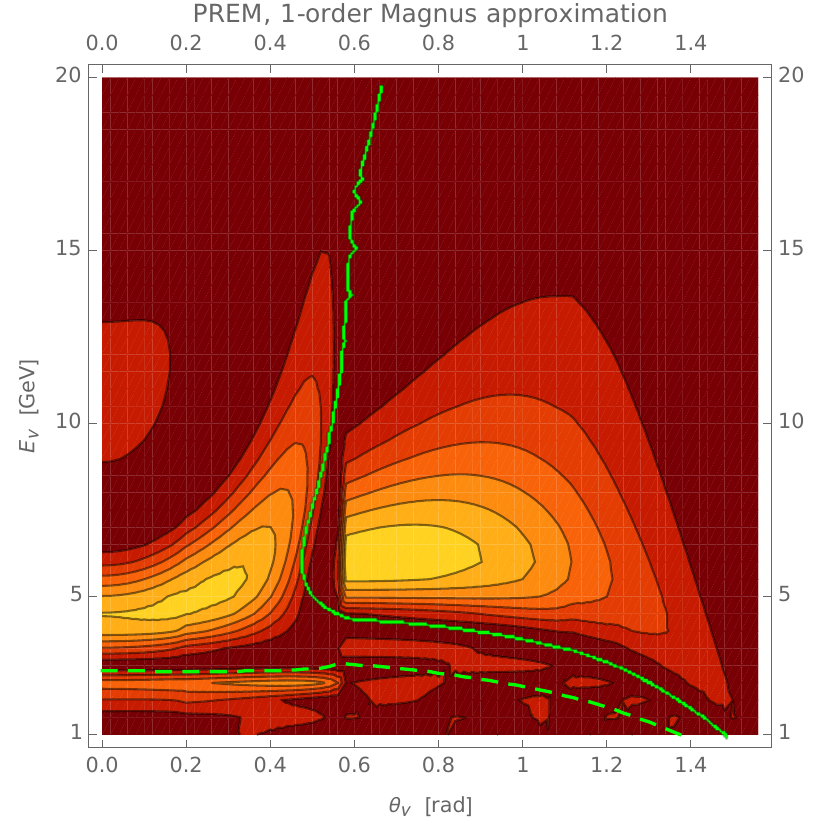}~
	\includegraphics[height=4.6cm]{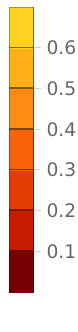}~
    \includegraphics[height=5.2cm]{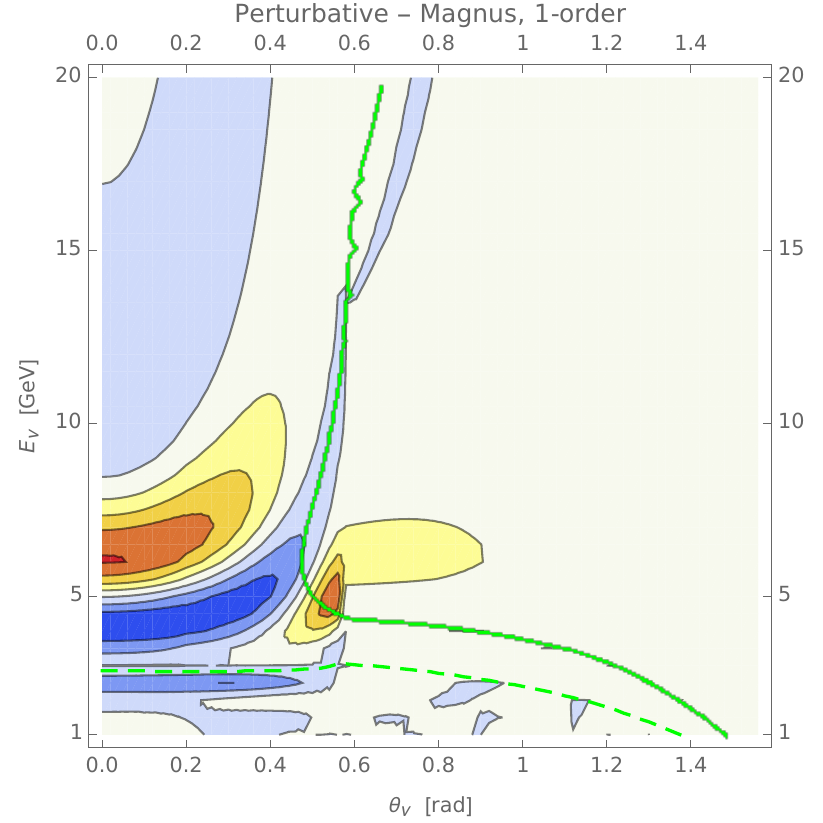}~
	\includegraphics[height=4.6cm]{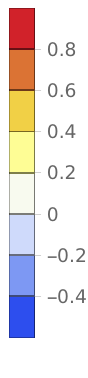}
	\caption{The oscillograms for PREM for neutrinos evolved in a single shell: the first-order Magnus iteration ({\it left}) and difference between the first-order perturbative and Magnus iterations ({\it right}). The solid and dashed green lines represent the convergence bound of Eq.~\eqref{eq:convergence} calculated for $x_c=L$ and $x_c=L/2$, correspondingly, where $L$ is the entire length of the neutrino path inside the Earth. Convergence of the Magnus series is guaranteed above the dashed line and in the upper-right relative to the solid one.
    }
	\label{Fig:oscillogram_PREM_1-order_Magnus}
\end{figure}

Next we show the probability obtained in the first-order Magnus approximation~\cite{Supanitsky:2008eq}
\begin{flalign}\label{eq:Magnus}
    P_{ea}(L) = \sin^2\left( \frac{\Delta m^2}{2E_\nu}\sin2\theta \int\limits_{L/2}^L dy \cos\left( 2\int\limits_{L/2}^y dx \,\omega(V(x)) \right) \right),
\end{flalign}
which is valid for a symmetric neutrino potential $V(x)=V(L-x)$, 
and compare it with the above perturbative result in Fig.~\ref{Fig:oscillogram_PREM_1-order_Magnus}. For the mantle only crossing trajectories the results of the two methods are close to each other, while the essential difference in results for the core-crossing trajectories is again not surprising. 

One of the advantages of the exponential Magnus expansion is its convergence if the following condition is satisfied~\cite{Blanes:2008xlr}
\begin{flalign}%\label{eq:convergence}
    \pi > I_c\equiv \int\limits_0^{x_c} ||H(x)||_2 dx,
\end{flalign}
where $x_c$ is the radius of convergence, and $||.||_2$ denotes the Euclidean norm. For the Hamiltonian given in Eq.~\eqref{eq:H} %Eqs.~\eqref{eq:H(x)} and \eqref{eq:H(x)_2}, 
the above integral reads
\begin{flalign}\label{eq:convergence}
    I_c(E_\nu,\theta_\nu) = \int\limits_0^{x_c(\theta_\nu)} \frac{\Delta m^2}{4E_\nu} \sqrt{\sin^2{2\theta} + \left(\cos{2\theta} - \frac{2E_\nu}{\Delta m^2}V(x)\right)^2} dx.
\end{flalign}
We show the convergence limits resulting from Eq.~\eqref{eq:convergence} by the green lines in Fig.~\ref{Fig:oscillogram_PREM_1-order_Magnus}. The solid and dashed lines were obtained for the integration along the total neutrino path and the half-path, respectively. The former is relevant if the neutrino state evolution is calculated using a single matrix $S(L)$ ($U(t_f,t_0)$ in the notations of Ref.~\cite{Supanitsky:2008eq}). The later is relevant if the evolution is obtained as $S^T(L/2)S(L/2)$, which is valid for a symmetric potential $V(x)=V(L-x)$. However, $V(x)\neq V(L/2-x)$ and therefore calculation of $S(L/2)$ is carried out for a non-symmetric neutrino potential. We used this approach for obtaining usual perturbative corrections since our expressions for the evolution matrix are valid for a generic form of $V(x)$, while Eq.~\eqref{eq:Magnus} can not be applied in this case. 

Besides the issue of large density jump between mantle and core, due to which the expansions in $\Delta V$ do not work properly, 
description of the neutrino state evolution inside the Earth by a single evolution matrix includes averaging of $V$ along either entire neutrino path or half-path in case of the symmetric potential. However, $\langle\Delta m_M^2\rangle$, $\langle\sin2\theta_M\rangle$ and $\Delta V$ are rather sensitive to the value of $\bar V$. The resulting approximation for $S(L)$ is very rough. 
More precise result can be obtained by considering evolution of the neutrino state separately in different shells, which do not contain large density jumps and possess their own $\bar V$ values. 
In the following we perform discussed multi-shell evolution in the perturbation theory.

\subsection{Multishell evolution}

Instead of obtaining the evolution matrix for entire neutrino trajectory at once, i.\,e. taking the Earth as a single shell, the planet can be considered with respect to evolution of the neutrino state as a structure of two or more spherical shells with coinciding centers. Here we distinguish the spherical layers of an Earth model and the spherical shells related to a chosen scheme of neutrino evolution. These shells can coincide with the Earth's density layers. Alternatively, either some shells can span multiple layers or, conversely, some layers can span multiple shells. However, calculation of the neutrino state evolution complicates with the increase of the number of shells. 

Generically, ignoring possible effects of violation of the closure property of the time-evolution operator, for trajectories crossing $k$ shells with relatively weakly varying densities factorization of the evolution matrix reads
\begin{flalign}\label{eq:S(L)}
    S(L) = S(\tilde L_1) \dots S(\tilde L_{k}) S(L_{k}) \dots S(L_1)
\end{flalign}
in terms of 
\begin{flalign}
    S(L_j)=S(x_j,x_{j-1}), \qquad S(\tilde L_j)=S(x_{2k-j+1},x_{2k-j})
\end{flalign}
where $x_i$ are the initial and final coordinates on the considered sections of neutrino trajectory, $L_j$ and $\tilde L_j$ denote the distances traveled by neutrino in each layer towards the deepest one and than towards the surface, respectively, and the subscripts refer to the index numbers of the shells, which are numbered from the outer one. Obviously, 
\begin{flalign}
    L= \sum_{j=1}^{k}(L_j+\tilde L_j).
\end{flalign}
For symmetric potential $V(x)=V(L-x)$ we have $L_j=\tilde L_j$, and Eq.~\eqref{eq:S(L)} simplifies to
\begin{flalign}\label{eq:MultiS}
    S(L) = S(L_1)^T \dots S(L_{k})^T S(L_{k}) \dots S(L_1),
\end{flalign}
where $L_j$ denotes the half-length of the trajectory in the $j$-th shell, %which results in 
$L=2\sum_{j=1}^{k}L_j$. 

In the following we consider two-shell evolution, which was elaborated in Ref.~\cite{Akhmedov:2006hb} and a scheme with ten evolution shells coinciding with the PREM density layers. In the next oscillograms that introduce perturbative corrections for the chosen multishell schemes we also show the respective convergence bounds resulting from Eq.~\eqref{eq:convergence} to illustrate how the dangerous regions with potential lack of convergence shrink as the number of shells increases.

\subsection{Two-shell evolution}
The perturbative approach discussed in section~\ref{sec:prturbative} can be applied for segments of the neutrino trajectory with relatively weakly varying density. However, realistic Earth density profile contains several jumps, in particular, the density almost doubles on transition from the mantle to the core, which results in the values of %perturbation parameter 
$|\Delta V/\bar V|$ up to 0.6. Variations in $\bar\omega$ and $\sin2\bar\theta$ are even more dramatic. Typically, their values for $\bar V$ averaged in the core, in the mantle or along full trajectory differ by a factor of several. For this reason, in this section we utilize a two-shell scheme considering evolution of the neutrino state separately in the mantle and in the core. Such consideration reduces $|\Delta V/\bar V|$ by nearly three times. The analogous effects related to smaller density jumps, which take place at transitions from the upper mantle to the lower one and from the outer core to the inner one, are ignored here, but we take them into account in the next section. 

In 2LEM the average densities along the neutrino paths in the mantle and the core simply equal to $\rho_{\rm M}$ and $\rho_{\rm C}$, respectively. However, this is not the case in more realistic Earth density models discussed in section~\ref{sec:density_profiles}.

In 3LEM the neutrino path length in the core and the lower mantle can be written as
\begin{flalign}%\label{eq:rho_aver3}
	L_{\rm C} = \left\{
	\begin{array}{lll}
		2\sqrt{R_{\rm C}^2-r_0^2} & ~{\rm for}~ & r_0<R_{\rm C}, \\
		0 & ~{\rm for}~ & R_{\rm C}<r_0<R_\oplus, 
	\end{array}  \right.
\end{flalign}
and
\begin{flalign}%\label{eq:rho_aver3}
	L_{\rm LM} = \left\{
	\begin{array}{lll}
		2\left(\sqrt{R_{\rm LM}^2-r_0^2}-\sqrt{R_{\rm C}^2-r_0^2}\right) & ~{\rm for}~ & r_0<R_{\rm C}, \\
		2\sqrt{R_{\rm LM}^2-r_0^2} & ~{\rm for}~ & R_{\rm C}<r_0<R_{\rm LM}, \\
		0 & ~{\rm for}~ & R_{\rm LM}<r_0<R_\oplus, 
	\end{array}  \right.
\end{flalign}
respectively. Then the average densities, corresponding to the neutrino paths in the core and mantle, read $\bar\rho_{\rm C} = \rho_C$ and 
\begin{flalign}
	\bar\rho_{\rm M} = \rho_{\rm LM}\frac{L_{\rm LM}-L_{\rm C}}{L-L_{\rm C}} + \rho_{\rm UM}\frac{L-L_{\rm LM}}{L-L_{\rm C}}.
\end{flalign}

\begin{figure}[tb]
	\centering
	\includegraphics[height=4cm]{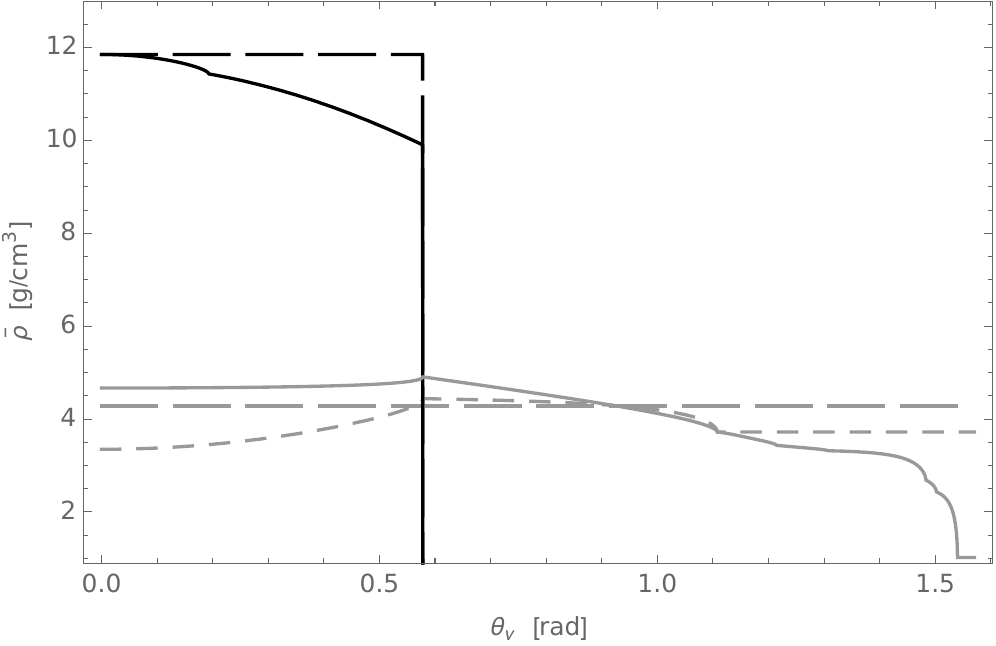}
    \includegraphics[height=4cm]{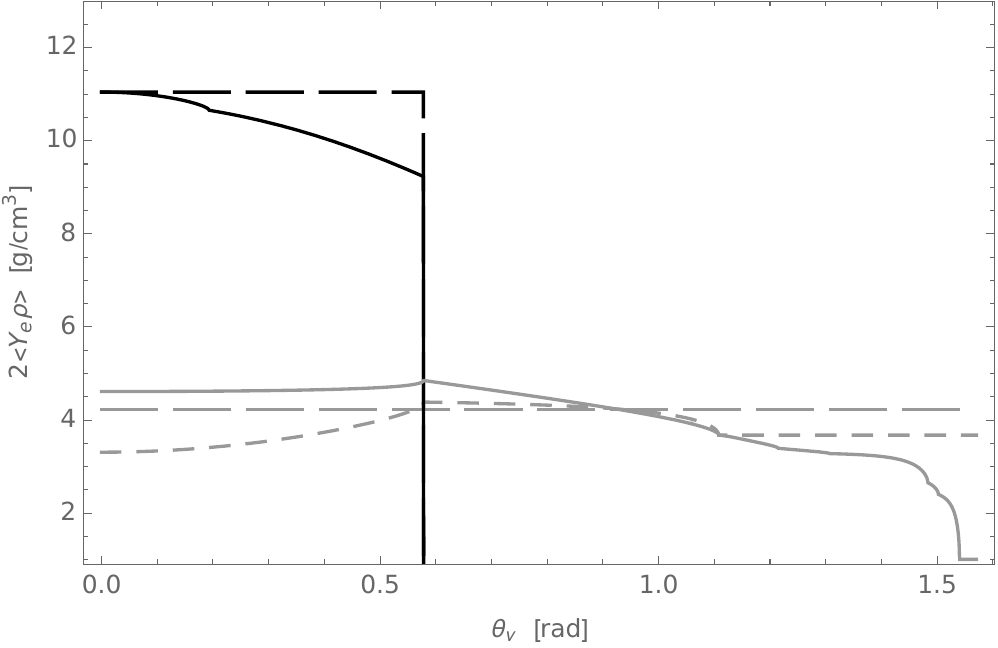}
	\caption{Average density $\bar\rho$ ({\it left}) and average product $\langle2Y_e\rho\rangle$ ({\it right}) 
    along the neutrino paths in the core (black lines) and mantle (gray lines) vs nadir angle $\theta_\nu$ for 2-, 3-layer and PREM models as represented by long-dashed, short-dashed and solid lines, respectively (for the paths inside the core the results of 2LEM and 3LEM coincide).}
	\label{Fig:rho_2-layer}
\end{figure}

In PREM, taking into account the density profile symmetry, $\rho(x)=\rho(L-x)$, the respective densities can be calculated as
\begin{flalign}\label{eq:rho_aver}
	\bar\rho_{\rm C}%(\theta_\nu)
    = \frac{1}{L_{\rm C}%(\theta_\nu)
    }\int\limits_{(L%(\theta_\nu)
    -L_{\rm C}%(\theta_\nu)
    )/2}^{(L%(\theta_\nu)
    +L_{\rm C}%(\theta_\nu)
    )/2}\rho(x%,\theta_\nu
    )dx, \qquad
%\end{flalign}
%\begin{flalign}\label{eq:rho_aver}
	\bar\rho_{\rm M} = \frac{2}{L_{\rm M}}\int\limits_0^{L_{\rm M}/2}\rho(x,\theta_\nu)dx,
\end{flalign}
where
\begin{flalign}
    L_{\rm M} = \left\{
	\begin{array}{lll}
		L - 2\sqrt{R_{\rm C}^2-r_0^2} & ~{\rm for}~ & r_0<R_{\rm C}, \\
		L & ~{\rm for}~ & R_{\rm C}<r_0<R_\oplus.
	\end{array}  \right.
\end{flalign}
The results for the average densities and the products $\frac{Y_e}{0.5} \rho$ that enter the neutrino potential are shown in Fig.~\ref{Fig:rho_2-layer} ({left}) and ({right}), respectively, for the discussed density schemes.

\begin{figure}[tb]
	\centering
	\includegraphics[height=5.4cm]{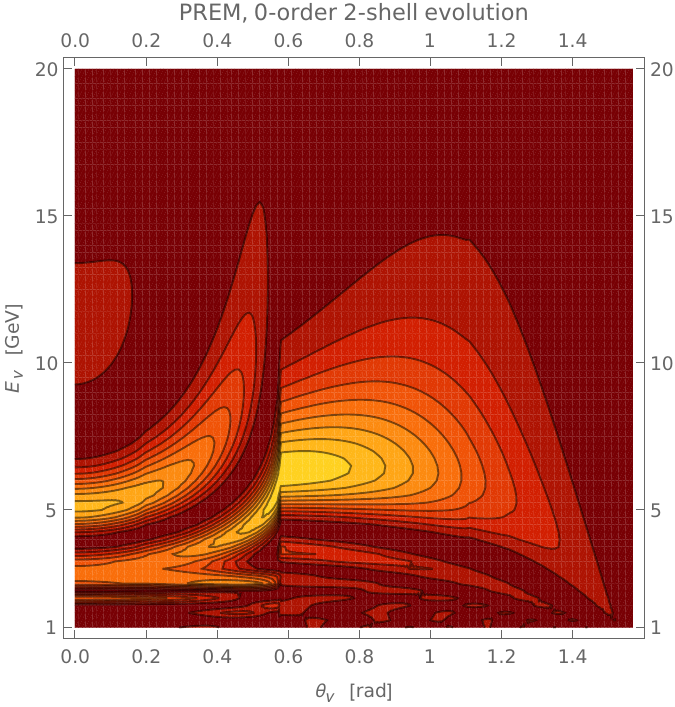}~
	\includegraphics[height=5cm]{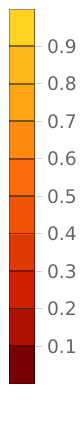}~~
    \includegraphics[height=5.4cm]{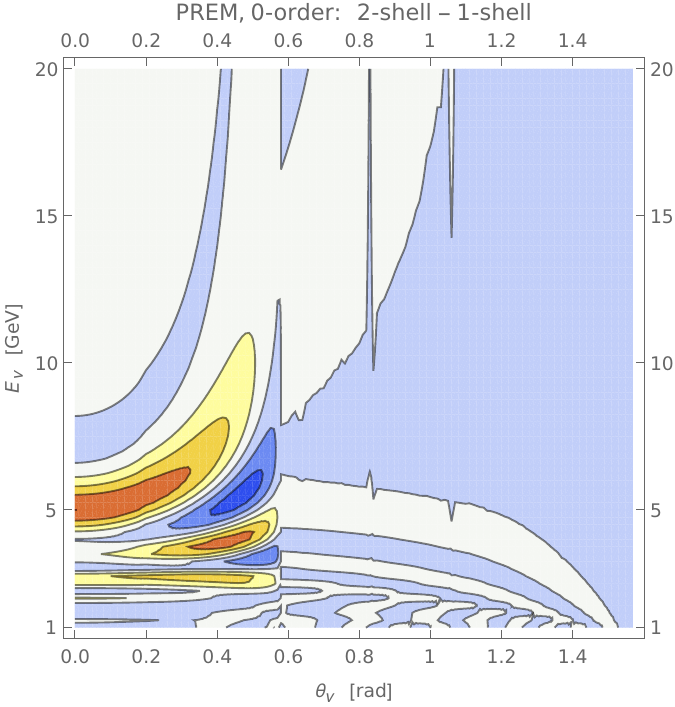}~
	\includegraphics[height=5cm]{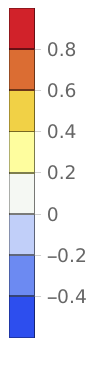}
    \caption{The zero-order oscillograms for PREM: with separate averaging in the mantle and in the core (two-shell evolution) ({\it left}), difference between two-shell and one-shell ({\it right}).}
	\label{Fig:oscillograms_0-order_two-shell}
\end{figure}

\begin{figure}[tb]
	\centering
	\includegraphics[height=5.4cm]{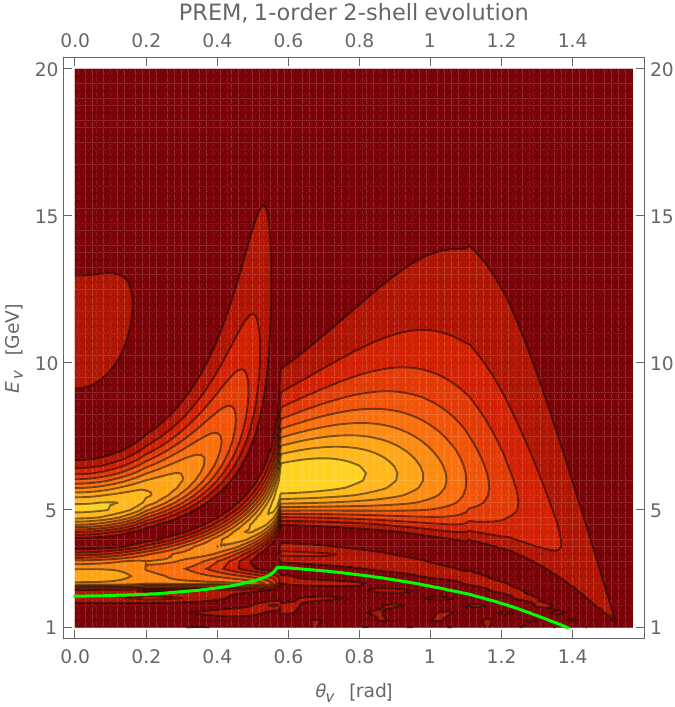}~
	\includegraphics[height=5cm]{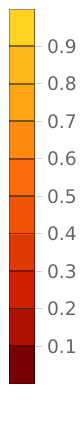}~~
    \includegraphics[height=5.4cm]{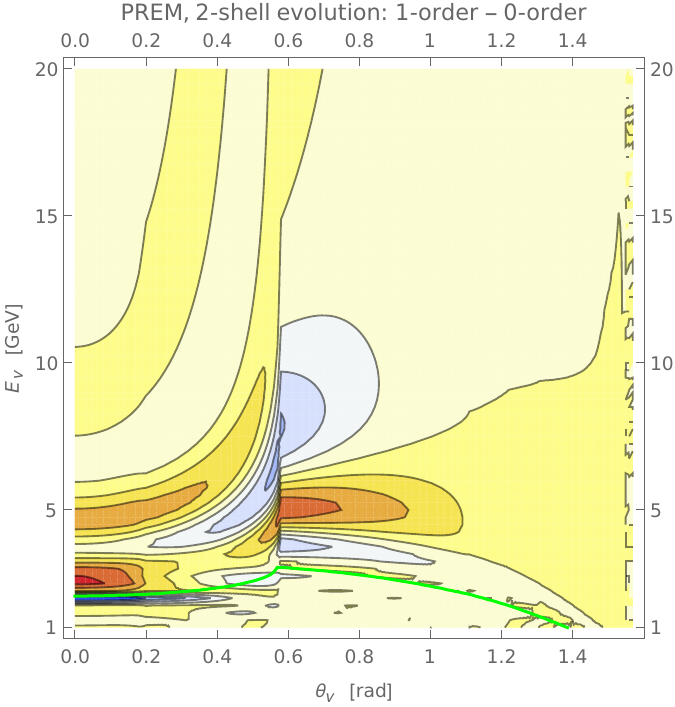}~
	\includegraphics[height=5cm]{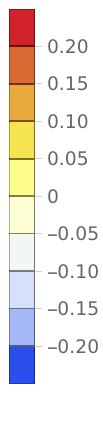}
	\caption{The oscillograms for PREM for neutrinos evolved in mantle and core separately: the first-order perturbative iteration ({\it left}) and difference between the first-order and zero-order iterations ({\it right}). The green lines represent the convergence lower bound for Magnus expansion.
    }
	\label{Fig:oscillogram_PREM_1-order_two-shell}
\end{figure}

The oscillogram for the zero-order perturbative evolution in the two shells, which is obtained using Eq.~\eqref{eq:Sbar}, is shown in Fig.~\ref{Fig:oscillograms_0-order_two-shell} (left). Its difference with the one-shell result is shown in Fig.~\ref{Fig:oscillograms_0-order_two-shell} (right). The first-order two-shell evolution result for PREM, which is unitarized according to Eqs.~\eqref{eq:S}, and its deviation from the zero-order iteration are shown in shown in Fig.~\ref{Fig:oscillogram_PREM_1-order_two-shell}. We remark that change in the oscillation probability from one-shell to two-shell evolution scheme is much more significant for the core-crossing trajectories with respect to the change from the zero-order to the first-order approximation. For this reason, we introduce more reliable 10-shell evolution scheme in the next section and only then calculate the second-order perturbative correction.

%%%%%%%%%%%%%%%%%%%%%%%%%%%%%%%%%%%%%%%%%%%%%%%%%%%%%%%%%%%%%%%%%%%
\subsection{Ten-shell evolution}

To obtain the neutrino oscillation probabilities for PREM with high precision we choose ten-shell evolution scheme, which shells correspond to the PREM layers with different density. Density variations inside each layer are small enough for validity and good results of perturbative expansion in $\Delta V$. The total evolution matrix we calculate according to Eq.~\eqref{eq:MultiS} with $k=10$ and all of the $S(L_k)$ we obtain for a selected order of the perturbation theory, except for the upper layers with constant densities for which the high-order corrections vanish. 

\begin{figure}[tb]
	\centering
	  \includegraphics[height=5.4cm]{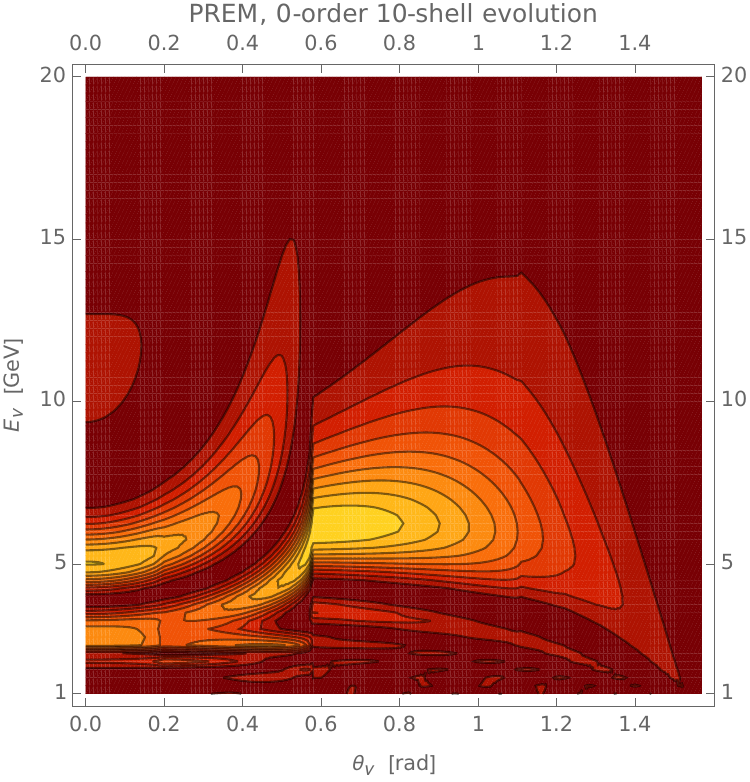}~
	\includegraphics[height=5cm]{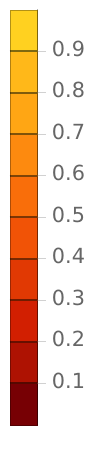}~~
    \includegraphics[height=5.4cm]{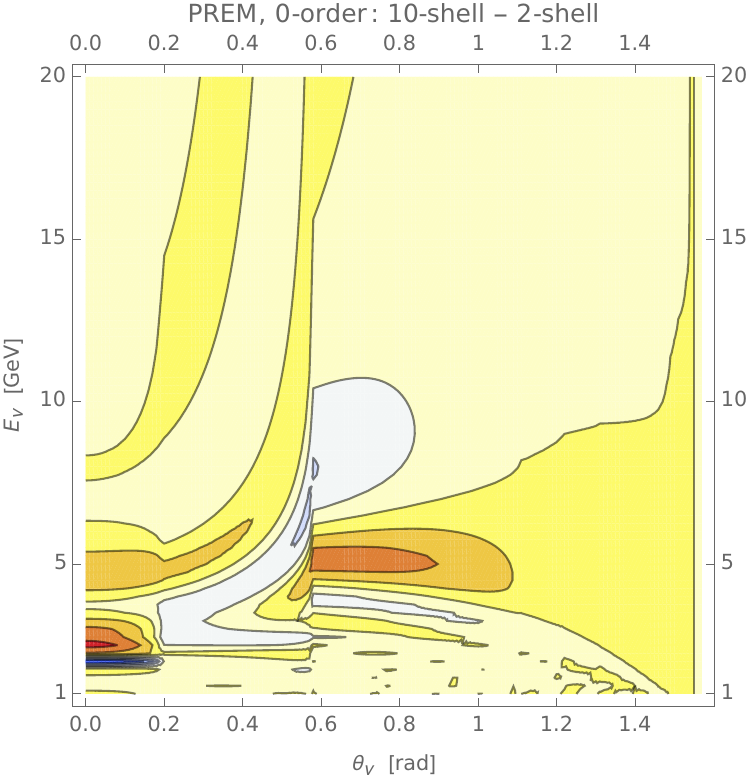}~
	\includegraphics[height=5cm]{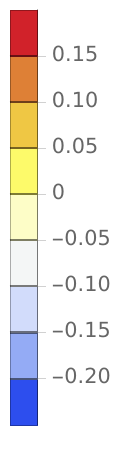}
	\caption{The zero-order oscillograms for PREM: 10-shell evolution ({\it left}), and difference between 10-shell and 2-shell evolution ({\it right}).}
	\label{Fig:oscillograms_0-order_ten-shell}
\end{figure}

\begin{figure}[tb]
	\centering
	\includegraphics[height=5.4cm]{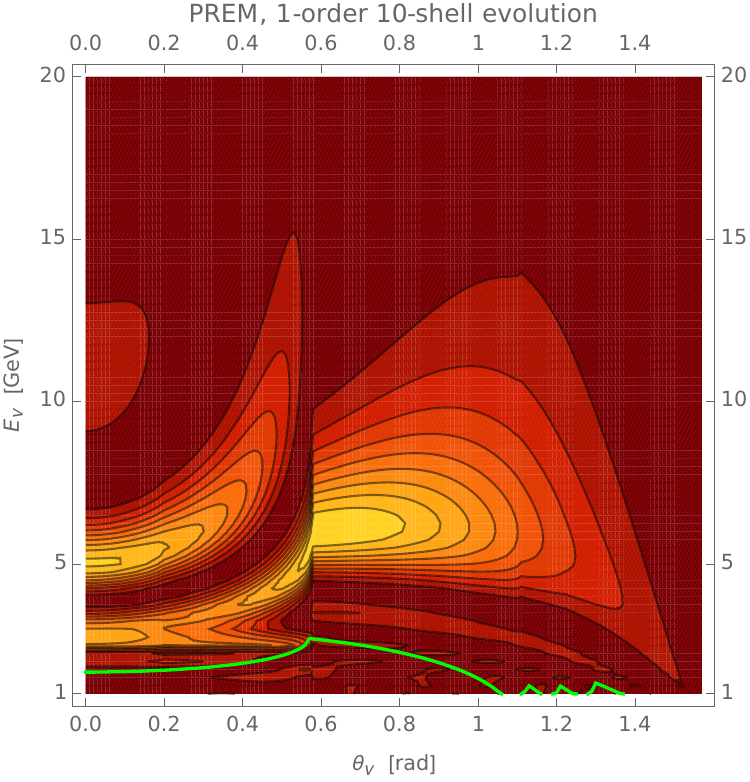}~
	\includegraphics[height=5cm]{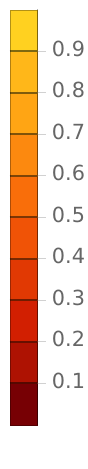}~~
    \includegraphics[height=5.4cm]{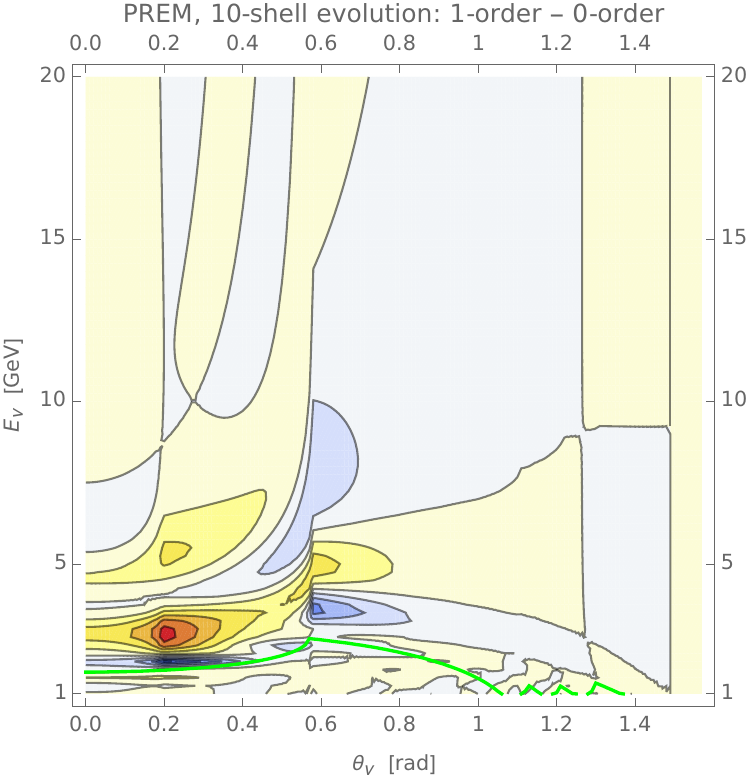}~
	\includegraphics[height=5cm]{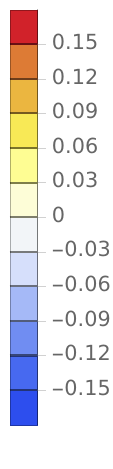}
	\caption{The oscillograms for PREM for neutrinos evolved in the ten shells separately: the first-order iteration ({\it left}) and difference between the first-order and zero-order iterations ({\it right}). The green lines represent the convergence bound for Magnus expansion.
    }
	\label{Fig:oscillogram_PREM_1-order_ten-shell}
\end{figure}

\begin{figure}[tb]
	\centering
	\includegraphics[height=5.4cm]{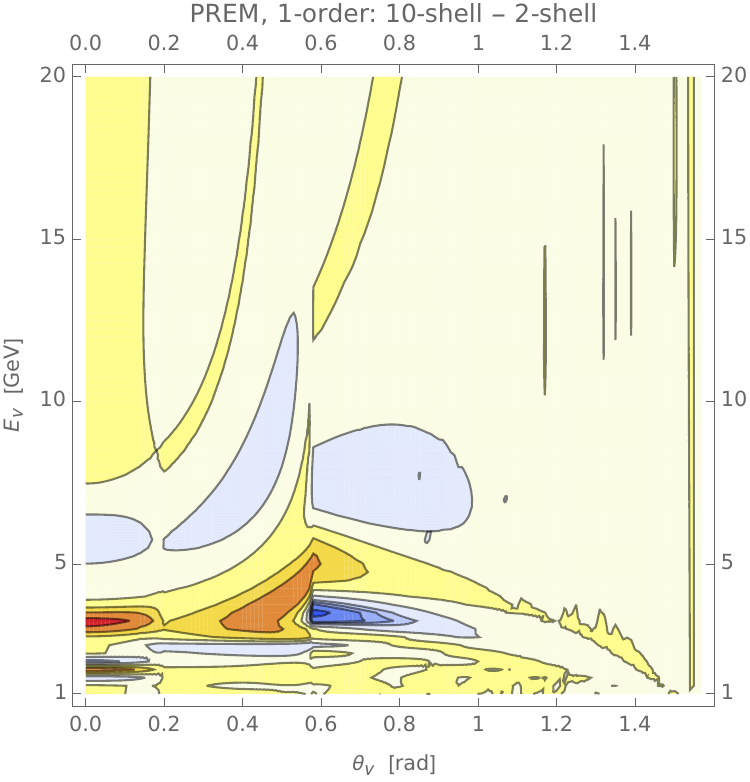}~
	\includegraphics[height=5cm]{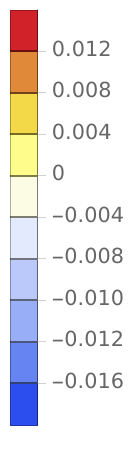}
	\caption{The oscillogram for PREM, which shows difference between neutrino evolutions in ten shells and in two shells, in the first-order iteration.% ({\it left}) and  ({\it right}).
    }
	\label{Fig:oscillogram_PREM_1-order_ten-shell-two-shell}
\end{figure}

\begin{figure}[tb]
	\centering
	\includegraphics[height=5.2cm]{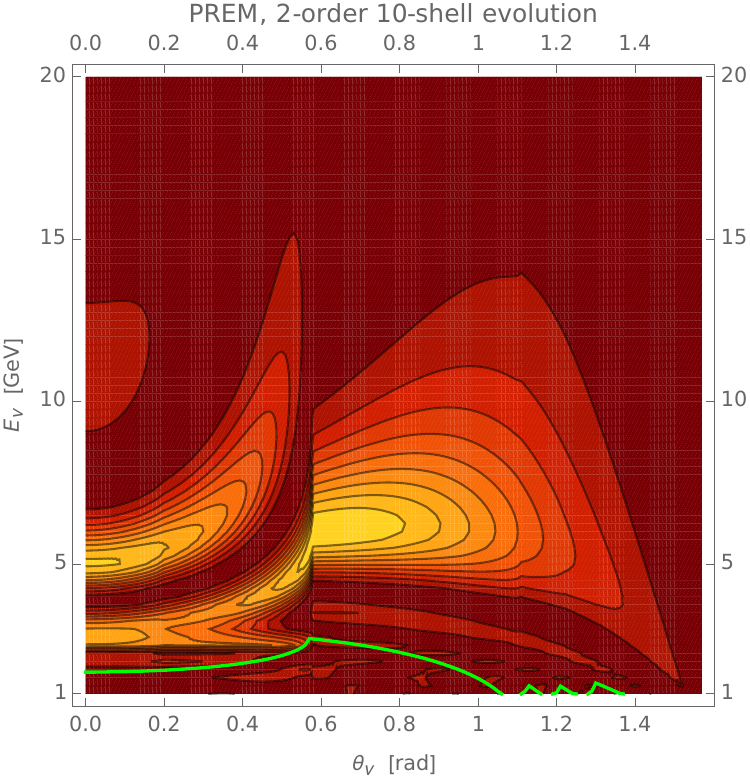}~
	\includegraphics[height=5cm]{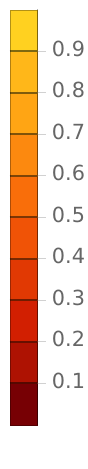}~~
    \includegraphics[height=5.2cm]{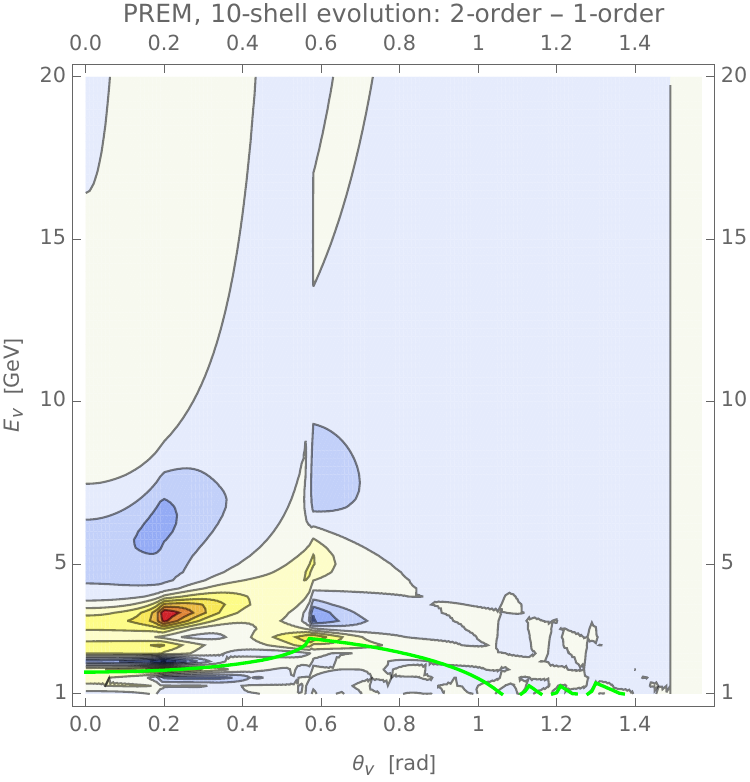}~
	\includegraphics[height=5cm]{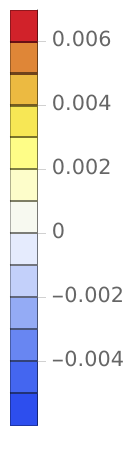}
	\caption{The oscillograms for PREM for neutrinos evolved in the ten shells separately: the second-order iteration ({\it left}) and difference between the second-order and first-order iterations ({\it right}). The green lines represent the convergence bound for Magnus expansion.
    }
	\label{Fig:oscillogram_PREM_2-order_ten-shell}
\end{figure}

The oscillogram for the zero-order ten-shell perturbative evolution and difference between the ten-shell and the two-shell results is shown in Fig.~\ref{Fig:oscillograms_0-order_ten-shell} (left) and (right), respectively. The unitarized first-order approximation and its deviation from the zero-order result are shown for the ten-shell evolution scheme in Fig.~\ref{Fig:oscillogram_PREM_1-order_ten-shell}. 
The effects of change from the two-shell to the ten-shell evolution at the zero order are comparable in size with the effects of change from the zero-order to the first-order iteration in the ten-shell scheme and exceed 15\% only in tiny regions of the parameter space. Within the first-order of perturbation theory change from the 2-shell to 10-shell evolution gives relatively small improvement, less than 2\%, which is seen from Fig.~\ref{Fig:oscillogram_PREM_1-order_ten-shell-two-shell}. However, the deviations of the second-order iteration from the first-order one in the ten-shell evolution scheme, which are presented in Fig.~\ref{Fig:oscillogram_PREM_2-order_ten-shell}~(right), are even several times smaller and do not exceed 0.7\%. These deviations approximately determine accuracy of the first-order iteration. Accuracy of the second-order ten-shell result, which is shown in Fig.~\ref{Fig:oscillogram_PREM_2-order_ten-shell}~(left), is likely several times better and can be determined using higher-order iterations.

\section{Discussion}\label{sec:discussion}

Our main result, the neutrino oscillogram obtained for PREM in the second-order of perturbation theory and for the ten-shell evolution scheme is shown in Fig.~\ref{Fig:oscillogram_PREM_2-order_ten-shell}~(left), where the green line represents the convergence limit for the Magnus exponential expansion. The area below this curve is dangerous, since in a region with a lack of convergence for the Magnus series usual perturbative series are also likely to diverge. To further improve convergence for neutrinos with energies less or about 2~GeV, larger number of the evolution shells can be used. In fact, wide shells can be divided into two or more subshells to reduce the appropriate upper limits of integration, according to Eqs.~\eqref{eq:convergence} and \eqref{eq:MultiS}. However, our next research interest concerns derivation of first- and second-order approximations for the Magnus expansion for non-symmetric neutrino potential with respect to the middle point of the neutrino trajectory. Later these expressions can be implemented to multishell evolution schemes to compare the results with the ones obtained in usual perturbative expansion, in particular, with the oscillograms presented in this paper, while convergence will be guaranteed for the areas above the discussed lower bounds. 

The obtained high precision neutrino oscillation probabilities can be combined with the theoretical unoscillated neutrino fluxes to generate predicted oscillated neutrino fluxes, using the known techniques~\cite{Jesus-Valls:2024tgd,Goos:2025rra}. The latter can be used for simulating the upcoming detector's signals and comparison with actual neutrino data, in particular, in the context of verification of seismic models of the Earth. However, this is a subject for next research.

\section{Conclusion}\label{sec:conclusion}

In this paper, we obtained the second-order perturbative correction to the neutrino oscillation probability for a generic relatively weakly varying neutrino potential. Then we generalized the known approach, which involves separate consideration of the neutrino state evolution in the Earth's mantle and core~\cite{Akhmedov:2006hb}, and introduced a ten-shell evolution scheme, which naturally suits for implementation 
of PREM with its density jumps. 
Using the above technique, we obtained the neutrino oscillograms of the Earth for arbitrary nadir angles and the range of neutrino energies of interest for NOT. In a common way, we neglected the solar neutrino mass splitting, which is a good approximation for considered neutrino energies. We compared the utilized approach with the Magnus exponential expansion and discussed its convergence limits. We demonstrated a reduction in the magnitude of perturbation corrections with increase of their order as well as achievement of greater accuracy when more sophisticated Earth models are used. We provided details of our analytical derivations and partly moved them to the Appendices for better readability. 

We hope that this work and forthcoming publications will help to improve the level of precision of the neutrino oscillation tomography of the Earth and answer the intriguing questions about the planet's deep interior.

\backmatter

%\bmhead{Supplementary information}
%
%If your article has accompanying supplementary file/s please state so here. 
%
%Authors reporting data from electrophoretic gels and blots should supply the full unprocessed scans for key as part of their Supplementary information. This may be requested by the editorial team/s if it is missing.
%
%Please refer to Journal-level guidance for any specific requirements.

%\bmhead{Acknowledgements}
%
%Acknowledgements are not compulsory. Where included they should be brief. Grant or contribution numbers may be acknowledged.

\begin{appendices}

\section{Iterative solution for the evolution equation}\label{sec:A1}

The solution of Eqs.~\eqref{eq:K}-\eqref{eq:Kj} reads 
\begin{flalign}
    K(x) &= \int_0^x dx_1A (x_1) -i \int_0^x dx_2 A(x_2) \int_0^{x_2} dx_1 A (x_1) \nonumber\\
	&+ (-i)^2 \int_0^x dx_3 A(x_3) \int_0^{x_3} dx_2 A (x_2)\int_0^{x_2} dx_1 A (x_1) +\dots 
\end{flalign}
By renaming the integration variables, we obtain
\begin{flalign}
    K(x) &= \int_0^x dx_1A (x_1) -i \int_0^x dx_1 A(x_1) \int_0^{x_1} dx_2 A (x_2) + \nonumber\\
	&+ (-i)^2 \int_0^x dx_1 A(x_1) \int_0^{x_1} dx_2 A (x_2)\int_0^{x_2} dx_3 A (x_3) +\dots 
\end{flalign}
Hence, 
\begin{flalign}
    \mathbb{I} -iK(x) &= \sum_{n=0}^\infty (-i)^n \int_0^{x} dx_1 \dots \int_0^{x_{n-1}} dx_n \mathcal{T} A (x_1)\dots A(x_n) \nonumber\\
	&= \sum_{n=0}^\infty \frac{(-i)^n}{n!} \int_0^{x} dx_1 \dots \int_0^{x} dx_n \mathcal{T} A (x_1)\dots A(x_n) = \mathcal{T}e^{-i\int\limits_0^{x} d\xi A(\xi)},
\end{flalign}
where the time-ordering operator $\mathcal{T}$ arranges a product of operators to the right from it in chronological order. 
Then final solution is given by Eq.~\eqref{eq:K_solution}.

\section{First-order solution}\label{sec:A2}

The right-hand side of Eq.~\eqref{eq:K1(x)} contains the adjoint action on $\sigma_3$, i.\,e. its rotation by an angle $2\phi$ along $\hat n$ axis:
\begin{flalign}
    \sigma_3 \to \bar S^\dag\sigma_3\bar S = R_{\hat n}(-2\phi)\sigma_3R_{\hat n}(2\phi),
\end{flalign}
where 
\begin{flalign}
    R_{\hat n}(\phi) =  \exp\left[{-i\frac{\phi}{2}\hat n\cdot{\vec\sigma}}\right]
\end{flalign}
is the rotation operator. 
Applying the Rodrigues' rotation formula, we obtain
\begin{flalign}\label{eq.4.5}
	R_{\hat n}(-2\phi)\sigma_3R_{\hat n}(2\phi) &= c_{2\phi} \sigma_3 + s_{2\phi} [\hat n\times\vec\sigma]_3 + (1-c_{2\phi}) n_3 (\hat n\cdot\vec\sigma) \nonumber\\
	&= - c_{2\bar\theta}s_{2\bar\theta}(1-c_{2\phi}) \sigma_1 + s_{2\bar\theta}s_{2\phi} \sigma_2 + (c_{2\phi}+c_{2\bar\theta}^2(1-c_{2\phi})) \sigma_3 \nonumber\\
	&= - c_{2\bar\theta}s_{2\bar\theta}(1-c_{2\phi}) \sigma_1 + s_{2\bar\theta}s_{2\phi} \sigma_2 + (s_{2\bar\theta}^2c_{2\phi}+ c_{2\bar\theta}^2) \sigma_3 \nonumber\\
    &= n_1 \left( c_{2\phi(x)}(n_1\sigma_3-n_3\sigma_1) + s_{2\phi(x)}\sigma_2 \right) + n_3 (\hat n\cdot\vec\sigma)
\end{flalign}
that agrees with the expression in the curly braces in Eq.~(4.5) of Ref.~\cite{Akhmedov:2006hb}. Due to Eq.~\eqref{eq:barV} $\langle \Delta V\rangle=0$. Hence, constant term $n_3 (\hat n\cdot\vec\sigma)$  
integrates out and we obtain
\begin{flalign}
	K_1(L) &=  \int_0^L \frac{\Delta V(x)}{2}\, e^{i\phi(x)\,\hat n\cdot{\vec\sigma}} \sigma_3 e^{-i\phi(x)\,\hat n\cdot{\vec\sigma}} dx \nonumber\\
	&= \frac{n_1}{2} \left( -[\hat n\times\vec\sigma]_2 \int_0^L \Delta V(x)c_{2\phi(x)} dx + \sigma_2 \int_0^L \Delta V(x)s_{2\phi(x)} dx \right) \label{eq.4.6}
\end{flalign}
that reproduces Eq.~(4.6) of Ref.~\cite{Akhmedov:2006hb}. 
For $\phi(x)=\bar\omega x$ in terms of the new variable $z = x - L/2$ we find
\begin{flalign}
	K_1(L) &= \int_{-L/2}^{L/2} \frac{\Delta V(z+\frac{L}{2})}{2} \, e^{i\bar\omega\,\hat n\cdot{\vec\sigma}(z+\frac{L}{2})} \sigma_3 e^{-i\bar\omega\,\hat n\cdot{\vec\sigma}(z+\frac{L}{2})} dz \nonumber\\
	&= e^{i\bar\omega\,\hat n\cdot{\vec\sigma}\frac{L}{2}} \left( \int_{-L/2}^{L/2} \frac{\Delta V(z+\frac{L}{2})}{2} \, e^{i\bar\omega\,\hat n\cdot{\vec\sigma}\,z} \sigma_3 e^{-i\bar\omega\,\hat n\cdot{\vec\sigma}\,z} dz \right) e^{-i\bar\omega\,\hat n\cdot{\vec\sigma}\frac{L}{2}} \nonumber\\
	&= \frac{n_1}{2}\, e^{i\bar\omega\,\hat n\cdot{\vec\sigma}\frac{L}{2}} \left( -[\hat n\times\vec\sigma]_2 {\rm Re}(I) + \sigma_2 {\rm Im}(I) 
    \right) e^{-i\bar\omega\,\hat n\cdot{\vec\sigma}\frac{L}{2}}, 
\end{flalign}
where in the last step we made the transformation similar to Eq.~(\ref{eq.4.6}) and used the notation of Eq.~\eqref{eq:I}. 
For the considered case with $n_2=0$ we obtain
\begin{flalign}\label{eq.4.7}
	\bar S(L) R_{\hat n}(-\bar\omega L) \left( -[\hat n\times\vec\sigma]_2 {\rm Re}(I)%\Delta I 
    + \sigma_2 {\rm Im}(I) 
    \right) R_{\hat n}(\bar\omega L) \nonumber\\
    =  -[\hat n\times\vec\sigma]_2 {\rm Re}(I) 
    + \sigma_2 {\rm Re}(I),
\end{flalign}
which follows from anticommutativity of different Pauli matrices and the expressions:
\begin{flalign}
    \{ \sigma_2, \hat n\cdot{\vec\sigma} \} = 0, \qquad\qquad  \{ [\hat n\times\vec\sigma]_2, \hat n\cdot{\vec\sigma} \} = 0.
\end{flalign}
Indeed, from $\{ X, Y \} = 0$ we have $[X, Y^{2k}] = 0$ and $\{ X, Y^{2k-1} \} = 0$ for $k\in\mathbb{N}$. Then
\begin{flalign}
    X e^Y &= X 
    \left( 1 + \sum_{k=1}^\infty \frac{Y^{2k-1}}{(2k-1)!} + \sum_{k=1}^\infty \frac{Y^{2k}}{(2k)!} \right) \nonumber\\
    &= \left( 1 - \sum_{k=1}^\infty \frac{Y^{2k-1}}{(2k-1)!} + \sum_{k=1}^\infty \frac{Y^{2k}}{(2k)!} \right) X = e^{-Y}X.
\end{flalign}
Hence, product of the exponential factors in Eq.~(\ref{eq.4.7}) gives unit matrix, and we end up with Eq.~\eqref{eq:DeltaS1}.

\end{appendices}


\bibliography{ZP_2026-bibliography}

\end{document}